\documentclass[preprint,pra,tightenlines,amsfonts,amssymb,%
showpacs,nofootinbib]{revtex4}
\usepackage{epsfig}
\usepackage{amsmath}
\usepackage{bm}

\renewcommand{\vec}[1]{\bm{\mathrm{#1}}}

\begin{document}

\title{Resonant $dd\mu$ Formation in Condensed Deuterium}

\author{Andrzej~Adamczak}%
 \email{andrzej.adamczak@ifj.edu.pl}
\affiliation{Institute of Nuclear Physics, Radzikowskiego 152, 
         PL-31-342 Krak\'ow, Poland}
\author{Mark~P.~Faifman}%
\email{faifman@imp.kiae.ru}
\affiliation{Russian Scientific Center, Kurchatov Institute,
RU-123182~Moscow, Russia}

\date{\today}

\begin{abstract}
  The rate of $dd\mu$ muonic molecule resonant formation in $d\mu$
  atom collision with a~condensed deuterium target is expressed in
  terms of a~single-particle response function. In particular, $dd\mu$
  formation in solid deuterium at low pressures is considered.
  Numerical calculations of the rate in the case of fcc
  polycrystalline deuterium at 3~K have been performed using the
  isotropic Debye model of solid. It is shown that the
  energy-dependent $dd\mu$ formation rates in the solid differ
  strongly from those obtained for D$_2$ gaseous targets, even at high
  $d\mu$ kinetic energies. Monte Carlo neutron spectra from $dd$
  fusion in $dd\mu$ molecules have been obtained for solid targets
  with different concentrations of ortho- and para-deuterium. The
  recent experimental results performed in low pressure solid targets
  (statistical mixture of ortho-D$_2$ and para-D$_2$) are explained by
  the presence of strong recoilless resonance peaks in the vicinity of
  2~meV and very slow deceleration of $d\mu$ atoms below 10~meV.
  A~good agreement between the calculated and experimental spectra is
  achieved when a~broadening of D$_2$ rovibrational levels in solid
  deuterium is taken into account. It has been shown that resonant
  $dd\mu$ formation with simultaneous phonon creation in solid gives
  only about 10\% contribution to the fusion neutron yield. The
  neutron time spectra calculated for pure ortho-D$_2$ and para-D$_2$
  targets are very similar. A~practically constant value of the mean
  $dd\mu$ formation rate, observed for different experimental
  conditions, is ascribed to the fact that all the recent measurements
  have been performed at temperatures $T\lesssim$~19~K, much lower
  than the target Debye temperature~$\Theta_{\text{D}}\approx$~110~K.
  In result, the formation rate, obtained in the limit
  $T/\Theta_{\text{D}}\ll$~1, depends weakly on the temperature.
\end{abstract}


\maketitle


\section{Introduction}
\label{sec:intro}

Theoretical study of resonant formation of the muonic molecule~$dd\mu$
in condensed deuterium targets is the main subject of this paper. The
resonant $dd\mu$ formation, first observed by Dzhelepov and
co-workers~\cite{dzhe66}, is a~key process of muon catalyzed fusion
($\mu$CF) in deuterium (see e.g.\ reviews~\cite{breu89,pono90}).
A~muonic deuterium atom $d\mu$ is created when a~negative
muon~$\mu^{-}$ is captured into an atomic orbital in a~deuterium
target. After $d\mu$ deexcitation to the $1S$ state and slowing down,
the $dd\mu$ molecule can be formed in $d\mu$ atom collision with one
of the D$_2$ target molecules. The resonant formation is possible due
to presence of a~loosely bound state of $dd\mu$, characterized by the
rotational number~$J=1$ and vibrational number~$v=1$, with binding
energy $|\varepsilon_{Jv=11}|\approx$~1.97~eV. This energy, according
to the Vesman mechanism~\cite{vesm67}, is completely transferred to
excited rovibrational states of the molecular complex $[(dd\mu)dee]$.
The scheme of calculation of $dd\mu$ formation rate in gaseous
deuterium has been developed for many
years~\cite{pono76,mens86,mens87,faif89}, and has lead to a~good
agreement with the experiments performed in gaseous
targets~\cite{scri93,peti99}. On the other hand, this theory, when
directly applied to solid deuterium targets, leads to strong
disagreement with the experimental
results~\cite{know96,demi96,know97}. Therefore, it is necessary to
calculate the $dd\mu$ formation rate with solid state effects taken
into account, which is the main purpose of this paper.

Our calculations are based on the theoretical results (transition
matrix elements, resonance energies) obtained in the case of $dd\mu$
formation in a~single D$_2$ molecule. In Sec.~\ref{sec:freemol} the
main formulas used for this case are briefly reported. A~general
formula for the energy-dependent $dd\mu$ formation rate in a~D$_2$
condensed target is derived in Sec.~\ref{sec:formalism}, using the Van
Hove formalism of the single-particle response function~\cite{vanh54}.
This formula is then applied (Sec.~\ref{sec:formsol}) for harmonic
solid targets, in particular for a~cubic Bravais lattice.  A~phonon
expansion of the response function is used to study phonon
contributions to the resonant formation. Numerical results for 3~K
zero pressure frozen deuterium targets (TRIUMF experimental
conditions~\cite{know96,know97}), with the fcc polycrystalline
structure, are shown in Sec.~\ref{sec:deut3K}.  The formation rates
have been calculated assuming the isotropic Debye model of the solid
and the values of Debye temperature and lattice constant observed in
neutron scattering experiments.

The calculated rates of resonant $dd\mu$ formation and back decay have
been used for Monte Carlo simulations of $dd$ fusion neutron and
proton time spectra. Since the initial distributions of $1S$ muonic
atom energy contain contributions from hot $d\mu$'s
($\sim$~1~eV)~\cite{mark94,abbo97}, influence of slow deceleration of
$d\mu$ atoms below 10~meV~\cite{adam96} on these time spectra is
investigated in Sec.~\ref{sec:Monte_Carlo_form}. The simulations take
into account the processes of incoherent and coherent $d\mu$ atom
scattering in solid deuterium. In particular, the Bragg scattering,
phonon scattering, and rovibrational transitions in D$_2$ molecules
are included. We consider a~dependence of the resonant formation rate
and time spectra on broadening of the rovibrational D$_2$ energy
levels, due to the binding of the molecules in the
lattice~\cite{vank83}.

Since it has been predicted in Refs.~\cite{filc96,mens96,guri99} that
strong $dd\mu$ formation takes place only in solid para-D$_2$, study
of this process in pure ortho-D$_2$ and para-D$_2$ targets is another
aim of this work. The neutron spectra calculated for these two solids
are discussed in Sec.~\ref{sec:Monte_Carlo_form}.


\section{Resonant formation in a~free molecule}
\label{sec:freemol}

First we consider resonant formation of the $dd\mu$ molecule in the
following reaction 
\begin{equation}
  \label{eq:reaction}
  (d\mu)_F + (\text{D}_2)^I_{\nu_iK_i} \to 
  \bigl[(dd\mu)_S^{Jv}dee\bigr]_{\nu_fK_f} \,,
\end{equation}
where D$_2$ is a~free deuterium molecule in the initial rovibrational
state $(\nu_iK_i)$ and the total nuclear spin~$\vec{I}$. The muonic
atom $d\mu$ has total spin $\vec{F}$ and CMS kinetic
energy~$\varepsilon$. The complex $[(dd\mu)dee]$ is created in the
rovibrational state~$(\nu_fK_f)$ and the molecular ion $dd\mu$, which
plays the role of a~heavy nucleus of the complex, has total
spin~$\vec S$.  The rate $\lambda^{SF}_{\nu_iK_i,\nu_fK_f}$ of the
process above depends on the elastic width
$\Gamma^{SF}_{\nu_fK_f,\nu_iK_i}$ of $[(dd\mu)dee]$ complex
decay~\cite{ostr80,lane83,gula85,padi88} in reactions
\begin{equation}
  \begin{split}
    \label{eq:ddm_channels}
    &\xrightarrow{\Gamma^{SF}_{\nu_fK_f,\nu_iK_i}}
    (d\mu)_F + (\text{D}_2)^I_{\nu_iK_i}    \\
    \bigl[(dd\mu)_S^{Jv}dee\bigr]_{\nu_fK_f}& \\
    &\xrightarrow[\tilde{\lambda}_f] 
    \quad \text{stabilization~processes,}
  \end{split}
\end{equation}
where $\tilde{\lambda}_f$ is the total rate of the stabilization
processes, i.e.\ deexcitation and nuclear fusion in~$dd\mu$
\begin{equation}
  \label{eq:dd_fusion}
  dd\mu \to 
  \begin{cases}
    \mu + t + p + 4.0 \text{ MeV} \\
    \mu +\, ^3\text{He} + n + 3.3 \text{ MeV} \\
    \mu^3\text{He} + n + 3.3 \text{ MeV} \,.
  \end{cases}
\end{equation}
When fusion takes place, the muon is generally released and can again
begin the $\mu$CF cycle. However, sometimes the muon is captured into
an atomic orbital of helium (sticking), which stops further reactions.

The value of $\Gamma^{SF}_{\nu_fK_f,\nu_iK_i}$ is given in atomic
units ($e=\hbar=m_e=$~1) by the formula
\begin{equation}
  \label{eq:el_width}
  \Gamma^{SF}_{\nu_fK_f,\nu_iK_i} = 2\pi A_{if} 
  \int \frac{d^3k}{(2\pi)^3} \, \lvert V_{if}(\varepsilon)\rvert^2
  \, \delta(\varepsilon_{if}-\varepsilon ) \,,
\end{equation}
where $V_{if}(\varepsilon)$ is a~transition matrix element and
$\varepsilon_{if}$ is a~resonance energy defined in
Ref.~\cite{faif89}. The factor~$A_{if}$ is due to averaging over
initial and summing over final projections of spins and angular
momenta of the system. Vector~$\vec{k}$ is the momentum of relative
$d\mu$ and D$_2$ motion
\begin{equation}
  \label{eq:momentum}
  \varepsilon = k^2/2{\mathcal M} \,,
\end{equation}
and ${\mathcal M}$ is the reduced mass of the system.   
Integration of Eq.~(\ref{eq:el_width}) over~$\vec{k}$ leads to
\begin{equation}
  \label{eq:gamma}
  \Gamma^{SF}_{\nu_fK_f,\nu_iK_i} = 
  \frac{{\mathcal M}\, k_{if}}{\pi} 
  A_{if}\lvert V_{if}(\varepsilon_{if})\rvert^2 \,,
  \qquad k_{if} = k(\varepsilon_{if}) \,.
\end{equation}
Since $\Gamma^{SF}_{\nu_fK_f,\nu_iK_i}$ and~$\tilde{\lambda}_f$ are
much lower ($\sim 10^{-3}$~meV) than $\varepsilon$, Vesman's model
can be applied and the energy-dependent resonant formation rate has
the Dirac delta function profile
\begin{equation}
  \label{eq:resrate}
\lambda^{SF}_{\nu_iK_i,\nu_fK_f}(\varepsilon)= 2\pi N
     B_{if}\bigl|V_{if}(\varepsilon)\bigr|^2 
    \delta(\varepsilon-\varepsilon_{if}) \,.    
\end{equation}
where $N$ is the density of deuterium nuclei in the target. According
to Ref.~\cite{faif89} the coefficients $A_{if}$ and $B_{if}$ in the
above equations are equal to
\begin{equation}
  \label{eq:AB_def}
  \begin{split}
    A_{if} & = 4\,  W_{SF}\, \xi(K_i) \,
    \frac{2K_i+1}{2K_f+1}  \,,  \\
    B_{if} & = 2\, W_{SF}\,
    \frac{2S+1}{2F+1} \,,
  \end{split}
\end{equation}
where
\begin{equation}
  \label{eq:Wigner}
  \begin{split}
    W_{SF} & = (2F+1)
    \begin{Bmatrix}
      \tfrac{1}{2} & 1 & F \\
                 1 & S & 1
    \end{Bmatrix}^{\!2}  \,, \\ 
    \xi(K_i) & = 
    \begin{cases}
      \frac{2}{3} & \text{ for } K_i=0 \,, \\
      \frac{1}{3} & \text{ for } K_i=1 \,,
    \end{cases}
  \end{split}
\end{equation}
and the curly brackets stand for the Wigner $6j$ symbol. In
formula~(\ref{eq:AB_def}) the usual Boltzmann factor describing the
population of rotational states in a~gas target is omitted because we
calculate the formation rate separately for each initial rotational
state. If the muonic atoms in a~gas have a~steady kinetic energy
distribution~$f(\varepsilon,T)$ at target temperature~$T$,
Eq.~(\ref{eq:resrate}) can be averaged over the atom motion leading to
a~mean resonant rate~$\lambda^{SF}_{\nu_iK_i,\nu_fK_f}(T)$.


\section{Resonant formation in a~condensed target}
\label{sec:formalism}

Since a~muonic deuterium atom can be approximately treated as a~small
neutron-like particle, methods used for description of neutron
scattering and absorption in condensed matter are applicable in the
case $dd\mu$ formation in dense deuterium targets. Below we adapt the
method developed by Lamb~\cite{lamb39}, and then generalized by Singwi
and Sj\"olander~\cite{sing60} using the Van Hove formalism of the
single-particle response function~${\mathcal S}_i$~\cite{vanh54}, for
calculation of the resonant $dd\mu$ formation rates.

\begin{figure}[htbp]
  \begin{center}
    \epsfig{file=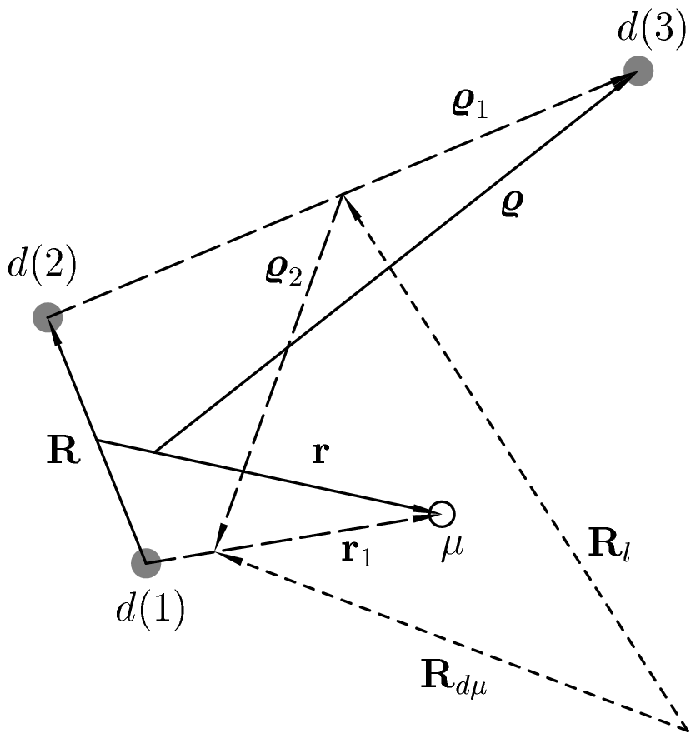, width=6.5cm}
    \caption{System of coordinates used for the calculation of
      resonant formation of the complex~$[(dd\mu)dee]$ in a~condensed
      deuterium target.}
    \label{fig:sys_cmplx}
  \end{center}
\end{figure}

A~Hamiltonian~$H_{\text{tot}}$ of a~system, consisting of a~$d\mu$
atom in the 1$S$ state and a~heavy condensed D$_2$ target, can be
written down as follows
\begin{equation}
  \label{eq:H_tot}
  H_ {\text{tot}} = \frac{1}{2M_{d\mu}}\nabla^2_{R_{d\mu}}
  + H_{d\mu}(\vec{r}_1) + H_{\text{D}_2}(\vec{\varrho}_1)
  + V(\vec{r}_1,\vec{\varrho}_1,\vec{\varrho}_2) + H  \,,
\end{equation}
where $M_{d\mu}$ is the $d\mu$ mass and $\vec{R}_{d\mu}$ denotes the
position of $d\mu$ center of mass in the coordinate frame connected
with the target (see Fig.~\ref{fig:sys_cmplx}). Operator $H_{d\mu}$ is
the Hamiltonian of a~free $d\mu$ atom, $\vec{r}_1$ is $d\mu$ internal
vector; $H_{\text{D}_2}$~denotes the internal Hamiltonian of a~free
D$_2$ molecule. It is assumed that $dd\mu$ formation takes place in
collision with the l-th D$_2$ target molecule.  The position of its
mass center in the target frame is denoted by~$\vec{R}_l$;
$\vec{\varrho}_1$~is a~vector connecting deuterons inside this
molecule. Function~$V$~stands for the potential of the $d\mu$--D$_2$
interaction~\cite{faif89}, leading to $dd\mu$ resonant formation.
Vector $\vec{\varrho}_2$ connects the $d\mu$ and D$_2$ centers of
mass. We neglect contributions to the potential~$V$ from the molecules
other than the l-th molecule because we assume here that distances
between different molecules in the target are much greater than the
D$_2$ size. The kinetic energy~$\varepsilon$ of the $d\mu$ atom and
its momentum~$\vec{k}$ in the target frame are connected by the
relation
\begin{equation}
  \label{eq:dm_kinetic}
  \varepsilon=k^2/2M_{d\mu} \,.
\end{equation}

\begin{figure}[htbp]
  \begin{center}
    \epsfig{file=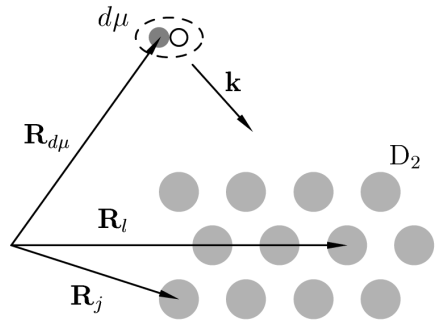, width=6cm}
    \caption{Position of impinging $d\mu$ atom with respect to the
      condensed target.}
    \label{fig:sys_lattice}
  \end{center}
\end{figure}

The Hamiltonian~$H$ of a~pure D$_2$ target, corresponding
to the initial target energy~$E_0$, has the form
\begin{equation}
  \label{eq:hamilt0}
  H = \sum_{j} \frac{1}{2M_{\text{mol}}}
  \nabla^2_{R_j} +\sum_{j}\sum_{j'\neq j} U_{jj'} \,,
\end{equation}
where $\vec{R}_j$ is the position of $j$-th molecule center of mass in
the target frame (Fig.~\ref{fig:sys_lattice}), $U_{jj'}$~denotes
interaction between the $j$-th and $j'$-th molecule, and
$M_{\text{mol}}$ is the mass of a~single target molecule.

The coordinate part~$\Psi_{\text{tot}}$ of the initial wave function
of the system can be written as a~product
\begin{equation}
  \label{eq:wf_tot1}
  \Psi_{\text{tot}} = \psi_{d\mu}^{1S}(\vec{r}_1)\, 
  \psi_{\text{D}_2}^{\nu_iK_i}(\vec{\varrho_1}) \,
  \exp(i\vec{k}\cdot\vec{R}_{d\mu})\, |0\rangle \,,
\end{equation}
where $|0\rangle$ stands for the initial wave function of the
condensed D$_2$ target, corresponding the total energy~$E_0$.
Eigenfunctions of the operators $H_{d\mu}$ and $H_{\text{D}_2}$ are
denoted by $\psi_{d\mu}^{1S}$ and $\psi_{\text{D}_2}^{\nu_iK_i}$,
respectively. Using the relation
$\vec{R}_{d\mu}=\vec{R}_l+\vec{\varrho}_2$, the wave function
$\Psi_{\text{tot}}$ takes the form
\begin{equation}
  \label{eq:wf_tot2}
  \Psi_{\text{tot}} = \psi_{d\mu}^{1S}(\vec{r}_1)\, 
  \psi_{\text{D}_2}^{\nu_iK_i}(\vec{\varrho_1}) \,
  \exp(i\vec{k}\cdot\vec{\varrho}_2)\, \exp(i\vec{k}\cdot\vec{R}_l) \,
  |0 \rangle \,,
\end{equation} 
which is similar to that used in the case of $dd\mu$ formation on
a~single D$_2$, except the factor 
$\exp(i\vec{k}\cdot\vec{R}_l)\,|0\rangle$. 
This factor depends only on positions of mass centers of the target 
molecules.

After formation of $[(dd\mu)dee]$ complex, the total Hamiltonian
of the system is well approximated by the operator~$H'_{\text{tot}}$
\begin{equation}
  \label{eq:H'_tot}
  H_{\text{tot}}\approx 
  H'_{\text{tot}} =  H_{dd\mu}(\vec{r},\vec{R}) 
  + H_{_C}(\vec{\varrho}) +V(\vec{\varrho},\vec{r},\vec{R}) 
  + \widetilde{H} \,,
\end{equation}
where $H_{dd\mu}$ is an internal Hamiltonian of $dd\mu$ molecular ion,
vectors $\vec{r}$ and $\vec{R}$ are its Jacobi coordinates. Relative
motion of $dd\mu$ and $d$ in the complex is described by
a~Hamiltonian~$H_{_C}$ which depends on the respective internal
vector~$\vec{\varrho}$. The final Hamiltonian $\widetilde{H}$ of the
target, with the eigenfunction~$|\widetilde{n}\rangle$ and energy
eigenvalue~$\widetilde{E}_n$, is expressed by the formula
\begin{equation}
  \label{eq:hamiltn}
  \begin{split}
    \widetilde{H} =&\frac{1}{2M_{C}}\nabla^2_{R_l} 
    +\sum_{j\neq l} \frac{1}{2M_{\text{mol}}}\nabla^2_{R_j}
    +\sum_{j}\sum_{j'\neq j} U_{jj'} \\
    =& -\left(1-\frac{M_{\text{mol}}}{M_C}\right)
    \frac{1}{2M_{\text{mol}}}\nabla^2_{R_l} + H
    =\Delta H + H \,,
  \end{split}
\end{equation} 
where $M_C$ is the mass of the complex. The respective coordinate
part~$\Psi'_{\text{tot}}$ of the total final wave function of the
system is
\begin{equation}
  \label{eq:wf'_tot}
  \Psi'_{\text{tot}}= \psi_{dd\mu}^{Jv}(\vec{r},\vec{R}) \, 
  \psi_{_C}^{\nu_fK_f}(\vec{\varrho}) \, |\widetilde{n}\rangle \,. 
\end{equation}
where $\psi_{dd\mu}^{Jv}$ and $\psi_{_C}^{\nu_fK_f}$ denote 
eigenfunctions of the Hamiltonians $H_{dd\mu}$ and~$H_{_C}$,
respectively.

The energy-dependent resonant $dd\mu$ formation
rate~$\lambda^{SF}_{\nu_iK_i,\nu_fK_f}(\varepsilon)$ in the condensed
target, for the initial $|0\rangle$ and final $|\widetilde{n}\rangle$
target states and a~fixed $d\mu$ total spin~$F$, is calculated using
the formula
\begin{equation}
  \label{eq:resratsol0}
  \lambda^{SF}_{\nu_iK_i,\nu_fK_f}(\varepsilon) = 
  2\pi N B_{if}\, \lvert {\mathcal A}_{i0,fn} \rvert^2 \, 
  \delta(\varepsilon-\varepsilon_{if}+E_0 -\widetilde{E}_n) \,,
\end{equation}
with the resonance condition 
\begin{equation}
  \label{eq:res_cond}
  \varepsilon + E_0 = \varepsilon_{if} + \widetilde{E}_n \,,
\end{equation}
taking into account the initial and final energy of the target. The
resonant energy for a~free D$_2$ is denoted by~$\varepsilon_{if}$ and
the transition matrix element is given by
\begin{equation}
  \label{eq:matrixel1}
  {\mathcal A}_{i0,fn} =  \langle \Psi'_{\text{tot}}\lvert  
  V \rvert \Psi_{\text{tot}} \rangle \,.
\end{equation}
Using Eqs. (\ref{eq:wf_tot2}) and
(\ref{eq:wf'_tot}) the matrix element~(\ref{eq:matrixel1}) can be
written as a~product 
\begin{equation}
  \label{eq:matrixel2}
  {\mathcal A}_{i0,fn} = \langle \widetilde{n}|\exp(i\vec{k}\cdot
  \vec{R}_l)|0\rangle \, V_{if}(\varepsilon) 
\end{equation}
where $V_{if}(\varepsilon)$ is the transition matrix element
calculated for a~single D$_2$ molecule~\cite{faif89}. The
rate~(\ref{eq:resratsol0}) can be additionally averaged over
a~distribution~$\rho_{n_0}$ of the initial target states at a~given
temperature~$T$ and summed over the final target states, which leads
to
\begin{equation}
  \begin{split}
  \label{eq:resratsol1}
  \lambda^{SF}_{\nu_iK_i,\nu_fK_f}(\varepsilon)= 
  2\pi N B_{if} |V_{if}(\varepsilon)|^2
  \sum_{n,n_0} &\rho_{n_0}\, \vert\langle \widetilde{n}\lvert
  \exp(i{\vec k}\cdot \vec{R}_l|0\rangle \rvert^2  \\ 
  &\times\delta(\varepsilon-\varepsilon_{if}+E_0
  -\widetilde{E}_n).
  \end{split}
\end{equation}
Factor $B_{if}$, defined by Eqs.~(\ref{eq:AB_def}), is due to the
averaging over the initial projections and summation over the final
projections of spin and rovibrational quantum numbers. This factor
takes also into account a~symmetrization of the total wave function of
$d\mu$+D$_2$ system over three deuterium nuclei.

Now we introduce a~time variable~$t$ to eliminate the
$\delta$~function in the equation above and then we involve
time-dependent operators, which is familiar in scattering theory (see,
e.g., Refs~\cite{akhi47,wick54}).  Using the Fourier expansion of the
$\delta$~function
\begin{equation}
  \label{eq:delta}
  \delta(\varepsilon-\varepsilon_{if}+E_0-\widetilde{E}_n)=
  \frac{1}{2\pi}\int_{-\infty}^{\infty}dt\, \exp\biglb(-it
  (\varepsilon-\varepsilon_{if}+E_0-\widetilde{E}_n)\bigrb)
\end{equation}
one has
\begin{equation}
  \begin{split}
  \label{eq:resratsol2}
  \lambda^{SF}_{\nu_iK_i,\nu_fK_f}(\varepsilon) =
  N B_{if}|V_{if}|^2 &\int_{-\infty}^{\infty} dt\, 
  \exp\biglb(-it\left(\varepsilon-\varepsilon_{if}\right)
  \bigrb)\sum_{n,n_0} \rho_{n_0} \\
  &\times\langle 0\vert\exp(-i\vec{k}\cdot\vec{R}_l)
  \vert\widetilde{n}\rangle 
  \langle\widetilde{n}\vert\exp(it\widetilde{E}_n)
  \exp(i\vec{k}\cdot\vec{R}_l)\exp(-itE_0)\vert 0\rangle \,.
  \end{split}
\end{equation}

Assuming that the perturbation operator~$\Delta H$ is
well-approximated by its mean value 
\begin{equation}
  \label{eq:res_shift}
  \Delta H \approx 
  \langle 0 \lvert \Delta H \rvert 0 \rangle \equiv
  \Delta\varepsilon_{if} =
  -\left(1-M_{\text{mol}}/M_C \right)\,{\mathcal E}_T < 0 \,,
\end{equation}
which is valid when the target relaxation time is much smaller than
the $dd\mu$ lifetime of the order of 10$^{-9}$~s, the matrix element
in Eq.~(\ref{eq:resratsol2}) can be expressed as
\begin{equation}
  \begin{split}
    \label{eq:matrix_Hei}
    &\langle\widetilde{n}\vert\exp(it\widetilde{E}_n)
    \exp(i\vec{k}\cdot\vec{R}_l)\exp(-itE_0)\vert 0\rangle \\
    &= \langle \widetilde{n}\vert\exp \biglb(it( H
      +\Delta H) \bigrb) \exp(i\vec{k}\cdot\vec{R}_l)
    \exp(-it H)\vert 0\rangle \\
    &\approx \langle \widetilde{n}\vert 
    \exp(it\Delta \varepsilon_{if})\, \exp(itH)
    \exp(i\vec{k}\cdot\vec{R}_l) \exp(-it H)\vert 0\rangle \\
    &= \langle \widetilde{n}\vert 
    \exp(it\Delta \varepsilon_{if})\,
    \exp\biglb(i\vec{k}\cdot\vec{R}_l(t)\bigrb)\vert 0\rangle \,,
  \end{split}
\end{equation}
where $\vec{R}_l(t)$ denotes the Heisenberg operator and 
${\mathcal E}_T$ in formula~(\ref{eq:res_shift}) is the mean kinetic 
energy of the target molecule at temperature~$T$.

Using the identity $\sum_n |\widetilde{n}\rangle\langle\widetilde{n}|=1$  
in Eq.~(\ref{eq:resratsol2}) we obtain 
\begin{equation}
  \label{eq:resratsol3}
  \begin{split}
    \lambda^{SF}_{\nu_iK_i,\nu_fK_f}(\varepsilon) = N B_{if}
    |V_{if}(\varepsilon)|^2 \int_{-\infty}^{\infty} & dt\,
    \exp\biglb(-it(\varepsilon-\varepsilon'_{if})\bigrb) \\
    &\times \bigl\langle \exp\biglb(
    -i\vec{k}\cdot\vec{R}_l(0)\bigrb) 
    \exp\biglb(i\vec{k}\cdot\vec{R}_l(t)\bigrb)\bigr\rangle_T \,,
  \end{split}
\end{equation}
where $\langle\cdots\rangle_T$ denotes both the quantum mechanical and
the statistical averaging at temperature~$T$, and 
$\varepsilon'_{if}$ being the resonance energy
\begin{equation}
  \label{eq:eres_sol}
  \varepsilon'_{if}=\varepsilon_{if}+
  \Delta\varepsilon_{if} \,, 
\end{equation}
shifted by $\Delta\varepsilon_{if}<0$. Note that such a~resonant
energy shift was neglected in papers~\cite{lamb39,sing60}, where
absorption of neutrons and $\gamma$-rays by heavy nuclei were
considered. An estimation of the shift in the case of
$\gamma$~emission from a~nucleus bound in a~solid, similar to
Eq.~(\ref{eq:res_shift}) was given in~Ref.~\cite{jose60}.

A~self pair correlation function~$G_s(\vec{r},t)$ is defined by
the following equation~\cite{vanh54}
\begin{equation}
  \label{eq:selfcor}
  \bigl\langle \exp\biglb( -i\vec{k}\cdot\vec{R}_l(0)\bigrb)
  \exp\biglb(i\vec{k}\cdot\vec{R}_l(t)\bigrb)\bigr\rangle_T =
  \int d^3r \,G_s(\vec{r},t)\,\exp(i\vec{k}\cdot\vec{r}) \,, 
\end{equation}
and the single-particle response
function~${\mathcal S}_i(\vec{\kappa},\omega)$ is given by the formula
\begin{equation}
  \label{eq:resp_def}
  {\mathcal S}_i(\vec{\kappa},\omega)=\frac{1}{2\pi}\int d^3r\,dt\;
  G_s(\vec{r},t)\,\exp\biglb( i(\vec{\kappa}\cdot
  \vec{r}-\omega t)\bigrb) \,. 
\end{equation}
Thus, by virtue of Eqs.~(\ref{eq:resratsol3}) and~(\ref{eq:resp_def}), the
resonant formation rate in a~condensed target can by expressed in
terms of the response function  
\begin{equation}
  \label{eq:resratsol4}
  \lambda^{SF}_{\nu_iK_i,\nu_fK_f}(\varepsilon) = 2\pi N B_{if}
  |V_{if}(\varepsilon)|^2\,{\mathcal S}_i(\vec{\kappa},\omega) \,,
\end{equation}
where the momentum transfer~$\vec{\kappa}$ and energy transfer~$\omega$
to the target are defined as follows
\begin{equation}
  \label{eq:omega_def}
  \vec{\kappa} = \vec{k} \,, \qquad
  \omega=\varepsilon-\varepsilon'_{if} \,.    
\end{equation}
The advantage of the Van Hove method is that all properties of the
target, for given momentum and energy transfers, are contained in the
factor~${\mathcal S}_i(\vec{k},\omega)$. It is possible to rigorously
calculate~${\mathcal S}_i$ in the case of a~perfect gas and in the case
of a~harmonic solid. However, a~liquid target or a~dense gas target 
is a~difficult problem to solve.

Proceeding as above one can obtain a~similar formula for 
$\Gamma^{SF'}_{\nu_fK_f,\nu_iK_i}$ in a~condensed target (in general,
$d\mu$ spin $F'$ after back decay can be different from $d\mu$ spin
$F$ before the formation)
\begin{equation}
  \label{eq:elwidthsol1}
  \begin{split}
    \Gamma^{SF'}_{\nu_fK_f,\nu_iK_i} & = 2\pi A_{if} 
    \int \frac{d^3k}{(2\pi)^3} \, \lvert V_{if}(\varepsilon)\rvert^2
    \widetilde{{\mathcal S}}_i(\vec{\kappa},\omega') \,, \\
    \omega' & = \tilde{\varepsilon}'_{if}- \varepsilon \,,
    \qquad \tilde{\varepsilon}'_{if} = 
    \varepsilon_{if} + \Delta 
    \tilde{\varepsilon}_{if} \,,
  \end{split}
\end{equation}
$\widetilde{{\mathcal S}}_i$ is the response function calculated
for the state~$\vert\widetilde{n}\rangle$ and
\begin{equation}
  \label{eq:eres_cplx}
  \Delta\tilde{\varepsilon}_{if} \equiv
  \langle \widetilde{n} \lvert \Delta H
  \rvert \widetilde{n} \rangle =  -\left(M_C / M_{\text{mol}}
    - 1\right)\, \widetilde{{\mathcal E}}_T \,,
\end{equation}
where $\widetilde{{\mathcal E}}_T$ denotes the mean kinetic energy
of the complex bound in the target.


\section{Resonant formation in a~harmonic solid}
\label{sec:formsol}

It has been shown by Van Hove~\cite{vanh54} that the self correlation
function in the case of a~gas or a~solid with cubic symmetry takes
the general form
\begin{equation}
  \label{eq:corr_gen}
   G_s(\vec{r},t) = \bigglb(\frac{M_{\text{mol}}}{ 2\pi\gamma(t)}
   \biggrb)^{3/2} \exp \bigglb(
   -\frac{M_{\text{mol}}}{2\gamma(t)}\, r^2 \biggrb) \,.
\end{equation}
For a~cubic Bravais lattice, in which each atom is at
a~center of inversion symmetry, $\gamma(t)$ is given by the formula  
\begin{equation}
  \label{eq:gamma_def}
  \gamma(t) =  \int_{-\infty}^{\infty} dw \, 
  \frac{Z(w)}{w} n_{_{\text{B}}}(w)\exp(-iwt) \, ,
\end{equation}
where $Z(w)$ is the normalized vibrational density of states such
that 
\begin{equation}
  \begin{split}
    \label{eq:z_def}
    \int_{0}^{\infty} dw \, Z(w) =1 \,, \quad
    & Z(w) = 0 \quad \text{ for } w > w_{\text{max}} \,, \\ 
    & Z(-w) \equiv Z(w) \,,
  \end{split}
\end{equation}
$n_{_{\text{B}}}(w)$ is the Bose factor
\begin{equation}
  \label{eq:Bose_fac}             
  n_{_{\text{B}}}(w) = \left[\exp(\beta w)-1\right]^{-1}\,, \quad
  \beta =(k_{\text{B}} T)^{-1}  \,.
\end{equation}
and the Boltzmann constant is denoted by~$k_{\text{B}}$. 

The response function~(\ref{eq:resp_def}), after substitution of 
Eqs.~(\ref{eq:corr_gen}),~(\ref{eq:gamma_def}) and integration
over~$\vec{r}$, can be written as follows  
\begin{equation}
  \label{eq:resp1}
  \begin{split}
    {\mathcal S}_i(\vec{\kappa},\omega) =&\frac{1}{2\pi}\exp\bigglb( 
    -\frac{\kappa^2}{2M_{\text{mol}}}\gamma(\infty)\biggrb) \\ 
    &\times \int_{-\infty}^{\infty} dt \, \exp(-i\omega t) 
    \exp\bigglb(\frac{\kappa^2}{2M_{\text{mol}}} \left[ 
    \gamma(\infty)-\gamma(t) \right] \biggrb) \,,
  \end{split}
\end{equation}
$\gamma(\infty)$ denotes the limit of~$\gamma(t)$ at~$t\to\infty$.
This formula can be expanded in a~power series of the momentum
transfer~$\kappa$, which leads to
\begin{equation}
  \label{eq:resp2}
    {\mathcal S}_i(\vec{\kappa},\omega) = \exp(-2W) \left[ 
    \delta(\omega) +\sum_{n=1}^{\infty} g_n(\omega,T) 
    \frac{(2W)^n}{n !} \right] \,, 
\end{equation}
where $2W$~is the Debye-Waller factor, familiar in the theory of
neutron scattering,
\begin{equation}
  \label{eq:2W}
  2W = \frac{\kappa^2}{2M_{\text{mol}}}\,\gamma(\infty) 
  = \frac{\kappa^2}{2M_{\text{mol}}} \int_0^{\infty} dw \,
  \frac{Z(w)}{w}\coth\left(\tfrac{1}{2} \beta w \right) \,,
\end{equation}
and the functions~$g_n$ are given by
\begin{equation}
  \label{eq:g_n}
  \begin{split}
    & g_1(w,T)  = \frac{1}{\gamma(\infty)} 
    \frac{Z(w)}{w} \left[n_{_{\text{B}}}(w)+1\right] \,, \\
    & g_n(w,T)  = \int_{-\infty}^{\infty} dw' \,
    g_1(w-w',T)\, g_{n-1}(w',T) \,, \\
    & \int_{-\infty}^{\infty} dw \, g_n(w) = 1 \,.
  \end{split}
\end{equation}
In the case of a~cubic crystal structure $2W$~can also be expressed as
\begin{equation}
  \label{eq:2W_u}
  2W =\tfrac{1}{3}\langle 0\lvert\vec{u}^2\rvert 0\rangle \kappa^2 \,,
\end{equation}
where $\vec{u}$ is the displacement of a~molecule from its lattice site.
Substitution of Eq.~(\ref{eq:resp2}) to Eq.~(\ref{eq:resratsol4})
leads to the following formation rate
\begin{equation}
  \label{eq:resratsol5}
  \lambda^{SF}_{\nu_iK_i,\nu_fK_f} (\varepsilon) = 2\pi N B_{if} 
  \lvert V_{if}(\varepsilon)\rvert^2 \exp(-2W) \left[ \delta(\omega) 
  +\sum_{n=1}^{\infty} g_n(\omega,T) \frac{(2W)^n}{n !} \right] \,, 
\end{equation}
The first term in expansion~(\ref{eq:resratsol5}) represents a~sharp
peak describing the $\delta$~profile recoilless formation. The next terms
give broad distributions corresponding to subsequent multi-phonon
processes. In particular, the term with~$n=1$ describes formation
connected with creation or annihilation of one phonon.

If $2W\ll 1$ we deal with so-called strong binding~\cite{lamb39} where
only the few lowest terms in the above expansion are important. On the
other hand, in the limit $2W\gg 1$ (weak-binding) many multi-phonon
terms give comparable contributions to~(\ref{eq:resratsol5}).
Therefore, for sufficiently large~$\kappa^2$ it is convenient to use
the impulse approximation in which $\gamma(t)$ is replaced by its
value near~$t=0$
\begin{equation}
  \label{eq:gam0}
  \gamma(t) \approx \gamma(0)+ it - \tfrac{2}{3}{\mathcal E}_T \,.
\end{equation}
This leads to the asymptotic formula for~${\mathcal S}_i$ 
\begin{equation}
  \label{eq:resp_asym}
  {\mathcal S}_i(\vec{\kappa},\omega) = \frac{1}{\Delta \sqrt{\pi}}
  \exp\bigglb(-\left(\frac{\omega-{\mathcal R}}{\Delta}\right)^2
  \biggrb)\,,
\end{equation}
where
\begin{equation}
  \label{eq:Dopp_width}
  \Delta = 2\sqrt{\tfrac{2}{3}{\mathcal E}_T {\mathcal R}} \,, 
  \qquad {\mathcal R} =\frac{\kappa^2}{2M_{\text{mol}}} \,.
\end{equation}
The mean kinetic energy~${\mathcal E}_T$ of a~molecule in the solid,
which also determines the resonance energy shift~(\ref{eq:res_shift}),
is equal to
\begin{equation}
  \label{eq:meankin}
  {\mathcal E}_T = \tfrac{3}{2}\int_0^{\infty} dw \,
  Z(w)\,w \left[n_{_{\text{B}}}(w)+\tfrac{1}{2}\right] \,.
\end{equation}
The energy ${\mathcal E}_T$ contains a~contribution from the
zero-point vibrations and it approaches $3k_{\text{B}}T/2$ only at
high temperatures $T\gg w_{\text{max}}/k_{\text{B}}$.
Function~(\ref{eq:resp_asym}) is a~Gaussian with response centered at
the recoil energy~${\mathcal R}$. Therefore in the weak binding region
the resonant formation rate takes the Doppler form obtained by Bethe
and Placzek~\footnote{In fact, formula~(\ref{eq:resp_asym}) is the
  limit of the Bethe formula in the case of a~very narrow natural
  resonance width~$\Gamma\to 0$.}  for resonant absorption of neutrons
in gas targets~\cite{beth37}.  However, the resonance
width~(\ref{eq:Dopp_width}) in the solid at temperature~$T$ is
different from the Doppler width in a~Maxwellian gas
$\Delta_{\text{gas}}=2\sqrt{k_{\text{B}} T {\mathcal R}}$ unless the
temperature is sufficiently high. This phenomenon was pointed out by
Lamb in his paper~\cite{lamb39} concerning resonant neutron absorption
in solid crystals. By virtue of the equations above one can introduce
for the solid an effective temperature~$T_{\text{eff}}$
\begin{equation}
  \label{eq:T_eff}
  T_{\text{eff}} = \tfrac{2}{3}\, {\mathcal E}_T / k_{\text{B}} \,.
\end{equation}


\section{Resonant formation in frozen deuterium}
\label{sec:deut3K}

The following considerations concern the solid deuterium crystals used
in~the TRIUMF experiments~\cite{mars93,know96b}, though the results
presented below can be applied to targets obtained in similar
conditions~\cite{demi96,stra96}. At TRIUMF thin solid deuterium layers
have been formed by rapid freezing of gaseous~D$_2$ on gold foils at
$T=3$~K and zero pressure. According to Ref.~\cite{silv80} such
deuterium layers have the face-centered cubic (fcc) polycrystalline
structure. Since the distance between the neighboring molecules is
a~few times greater than the diameter of a~D$_2$ molecule and the Van
der Waals force that binds the solid is weak, one can neglect
perturbations of the resonant formation potential~$V$ due to these
neighbors.

The deuterium crystals at zero pressure are quantum molecular
crystals. The amplitude of zero-point vibration at 3~K equals 15\% of
the nearest neighbor distance. A~single-particle potential in this
case is not harmonic and the standard lattice dynamics leads to
imaginary phonon frequencies. However, the standard dynamics can be
applied after a~renormalization of the interaction potential, taking
into account the short-range pair correlations between movement of the
neighbors~\cite{silv80}. In result, the theoretical
calculations~\cite{klei70} of the phonon dispersion relations give
a~good agreement with the neutron scattering experiments~\cite{niel73}
and the Debye model for solid deuterium can be used as a~good
approximation of the phonon energy distribution
\begin{equation}
  \label{eq:Z}
  Z(w)=
  \begin{cases}
    3\, w^2/w_{_{\text{D}}}^3 & \text{ if } \,
    w \leq w_{_{\text{D}}} \,, \\
    0 & \text{ if } \, w > w_{_{\text{D}}} \,,
  \end{cases}
\end{equation}
with the Debye energy~$w_{_{\text{D}}}=k_{\text{B}}\Theta_{\text{D}}$
and Debye temperature~$\Theta_{\text{D}}$ taken from the neutron
experiments.  For $T=3$~K we use the Debye model of an isotropic solid
with $\Theta_{\text{D}}=108$~K corresponding to the maximal phonon
energy~$w_{_{\text{D}}}=9.3$~meV.  Thus, we are dealing with the limit
$T/\Theta_{\text{D}}\ll 1$ where
\begin{equation}
  \label{eq:2W_asym}
  \gamma(\infty) = \tfrac{3}{2}\, w_{_{\text{D}}}^{-1} \,, \quad 
  {\mathcal E}_T = \tfrac{9}{16}\, w_{_{\text{D}}} 
  \approx \text{5.2~meV}, \quad
    T_{\text{eff}} = \tfrac{3}{8}\, \Theta_{\text{D}} \approx
  \text{40~K} \,,
\end{equation}
are very good approximations of Eqs.~(\ref{eq:2W}), (\ref{eq:meankin})
and~(\ref{eq:T_eff}). The Debye-Waller factor and mean kinetic
energy~${\mathcal E}_T$ at lowest temperatures are determined by
contributions from the zero-point D$_2$ vibration in the lattice, and
therefore these quantities do not tend to zero at $T\to 0$. The
zero-point energy is not accessible energy, but its effects are always
present.

The values of the resonance energies depend on initial and final
rovibrational quantum numbers of the system. In solid hydrogens at low
pressures these quantum numbers remain good quantum numbers, but
excited energy levels broaden into energy bands (rotons and vibrons)
due to coupling between neighboring molecules~\cite{vank83}. The
calculations presented in the literature concern pure solid H$_2$, HD
and D$_2$ targets and only lowest quantum numbers. The problem of
a~heavier impurity, such as $(dd\mu)d$ complex in D$_2$, has not been
considered yet. However, knowing that the width of the rotational
bands can reach about 1~meV~\cite{vank83}, a~possible influence of
this effect on the calculated formation rates and fusion neutron time
spectra is discussed in the next section.

At low temperatures all D$_2$ molecules are in the ground vibrational
state $\nu_i=0$ and $dd\mu$ is formed via the excitation of the
complex to the state $\nu_f=7$. Unless a~catalyst is applied, rapidly
frozen deuterium is a~mixture of ortho-D$_2$ ($K_i=0$) and para-D$_2$
($K_i=1$). In the TRIUMF experiments gaseous deuterium was pumped
through a~hot palladium filter before freezing. Therefore the solid
target was a~statistical mixture (2:1) of the ortho- and para-states
($K_i=stat$). Since the para-ortho relaxation without a~catalyst is
very slow (0.06\%/h) in solid deuterium~\cite{soue86}, the population
of these states is not changed during experiments of a~few days.

\begin{figure}[htbp] 
  \begin{center}
    \epsfig{file=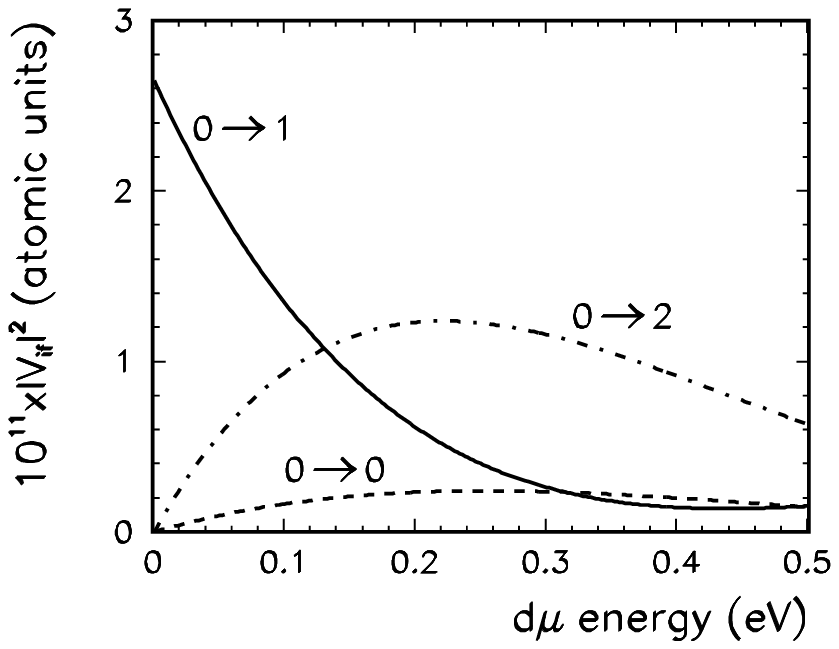, width=8.5cm}
    \caption{Transition matrix elements~$|V_{if}(\varepsilon)|^2$ for
      $K_i=0$ and $K_f=0,1,2$ versus $d\mu$ energy}
    \label{fig:v0f}
  \end{center}
\end{figure}

\begin{figure}[htbp] 
  \begin{center}
    \epsfig{file=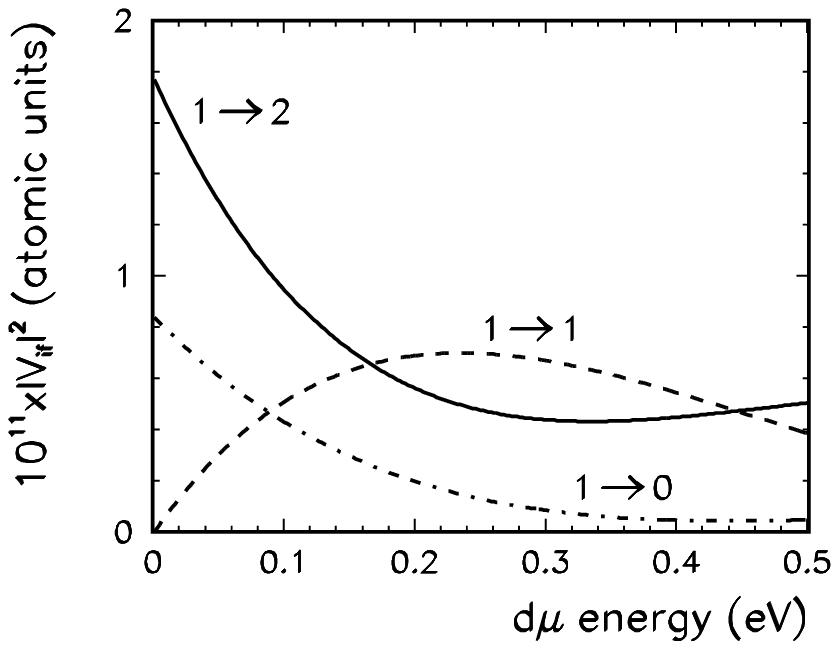, width=8.5cm}
    \caption{Transition matrix elements~$|V_{if}(\varepsilon)|^2$ for
      $K_i=1$ and $K_f=0,1,2$ versus $d\mu$ energy}  
    \label{fig:v1f}
  \end{center}%
\end{figure}

The lowest resonance energies~$\varepsilon_{if}$ and
$\varepsilon_{if}'$, for fixed $\nu_i$, $\nu_f$ and different values
of $F$, $K_i$, $S$ and $K_f$ are shown in
Table~\ref{tab:eres}~\cite{peti99}. A~few of them have negative
values, which means that to satisfy the resonance condition
$\varepsilon=\varepsilon_{if}$ an energy excess in the $d\mu$+D$_2$
system should be transferred to external degrees of freedom. This is
possible in dense targets, where energy of neighboring molecules can
be increased. Such an effect, due to triple collisions in gas targets,
has been firstly discussed in Ref.~\cite{mens86b}.  In a~solid, the
energy excess is lost through incoherent phonon creation. According
to~(\ref{eq:res_shift}), (\ref{eq:eres_sol}), and~(\ref{eq:2W_asym}),
in the considered 3~K solid deuterium all resonant energies
$\varepsilon_{if}'$ are shifted by 
$\Delta\varepsilon_{if}\approx -1.81$~meV. One can see that all
resonances for $F=\tfrac{1}{2}$ are placed at higher energies, which
is caused by $d\mu$ hyperfine splitting 
$\Delta E^{\text{hfs}}=$~48.5~meV. All resonance energies 
$\varepsilon_{if}'\lesssim w_{_{\text{D}}}\approx$~10~meV are connected 
with formation from the upper spin state $F=\tfrac{3}{2}$ of $d\mu$.  
However, only resonances corresponding to the dipole transitions
$K_i=0\to K_f=1$ and $K_i=1\to K_f=0,2$ can give a~significant
contribution to the formation rate at lowest energies.  Other
transition matrix elements described in Ref.~\cite{faif96} tend to
zero when $\varepsilon \to 0$ (see Figs.~\ref{fig:v0f}
and~\ref{fig:v1f} obtained for $K_i=0$ and $K_i=1$ ).

\begin{figure}[htbp]
  \begin{center}
    \epsfig{file=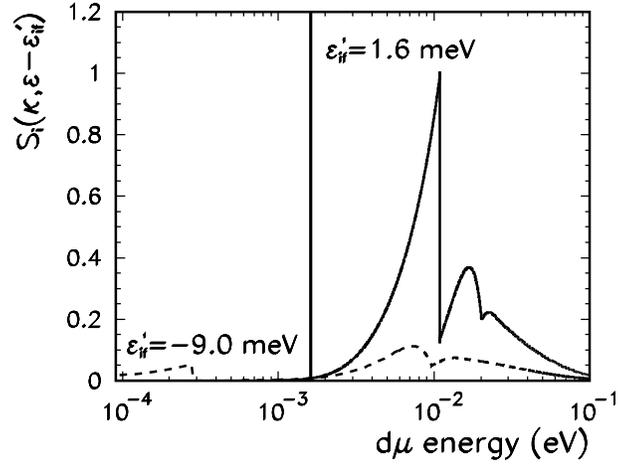, width=8.5cm}
    \caption{Response function
      ${\mathcal S}_i(\vec{\kappa},\varepsilon-\varepsilon_{if}')$ 
      (in arbitrary units) for the para-D$_2$ crystal at 3~K. The 
      dashed line is obtained for the subthreshold resonance
      $\varepsilon_{if}'\approx -9.0$~meV, the solid line corresponds
      to $\varepsilon_{if}'\approx 1.6$~meV. The vertical line
      represents the rigid lattice term
      $\delta(\varepsilon-\varepsilon_{if}')\exp(-2W)$.}
    \label{fig:resp}
  \end{center}
\end{figure}

The low energy rates ($\varepsilon\lesssim w_D$) are calculated using
formula~(\ref{eq:resratsol5}) with a~few most significant terms of the
response function expansion~(\ref{eq:resp2}) taken into account.
Fig.~\ref{fig:resp} shows the function 
${\mathcal S}_i(\vec{\kappa},\varepsilon-\varepsilon_{if}')$
corresponding to the two dipole transitions in para-D$_2$. The
subthreshold resonance, with $\varepsilon_{if}'\approx -9.0$~meV,
gives contributions to the formation rate only through the phonon
creation processes. For $\varepsilon_{if}'\approx 1.6$~meV, the
non-phonon process is possible and it is represented by a~vertical
line. Different peaks in this figure describe processes connected with
different numbers of created phonons. In particular, one-phonon
processes, which are proportional to $Z(w)$ with the characteristic
Debye cutoff, can be clearly distinguished. Since the $n$-phonon term
in~(\ref{eq:resp2}) is proportional to~$\kappa^{2n}$, the $dd\mu$
formation rate tends to zero at $\varepsilon \to 0$. Note that the
phonon annihilation gives negligible contribution to the rate at very
low target temperatures $T\ll\Theta_{\text{D}}$.

\begin{figure}[htbp]
  \begin{center}
    \epsfig{file=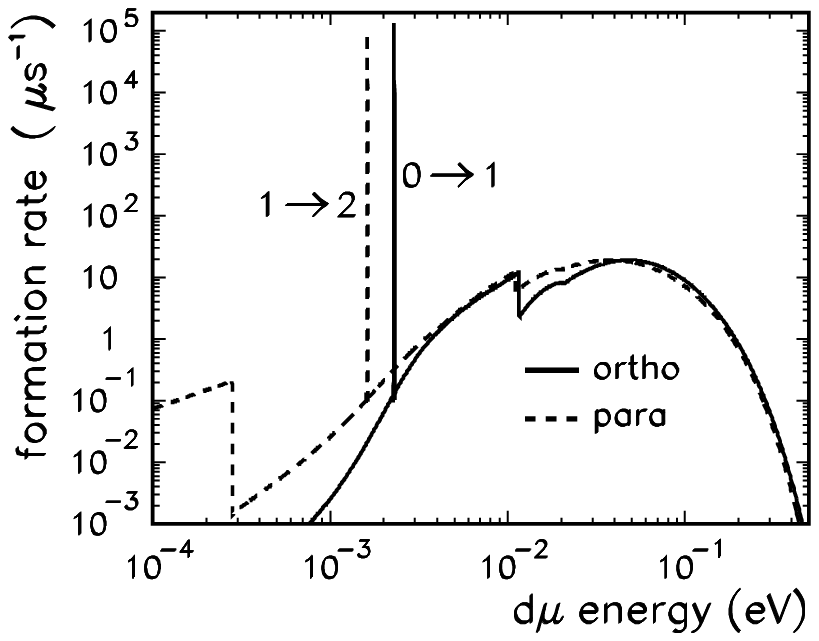, width=8.5cm}
    \caption{Formation rate $\lambda^F_{K_i}(\varepsilon)$ for
      $F=\tfrac{3}{2}$ in 3~K ortho-D$_2$ (solid line) and para-D$_2$ 
      (dashed line). The labels ``$1\to 2$'' and ``$0\to 1$'' denote 
      the rotational transition $K_i\to K_f$ corresponding to the 
      lowest non-phonon processes.}
    \label{fig:rate3_abs}
  \end{center}
\end{figure}

In order to compare the calculated formation rates with experiments 
the summed rates $\lambda^F_{K_i}(\varepsilon)$ are introduced 
\begin{equation}
  \label{eq:rate_sum}
  \lambda^F_{K_i}(\varepsilon) =  \sum_{K_f, S} 
  \lambda^{SF}_{\nu_i K_i \nu_f K_f} \,, \qquad \nu_i=0, 
  \quad \nu_f=7 \,.
\end{equation}
In Fig.~\ref{fig:rate3_abs} the formation rates
$\lambda^F_{K_i}(\varepsilon)$ in the solid ortho-D$_2$ and para-D$_2$
are shown for $F=\tfrac{3}{2}$. In the case of resonances satisfying
the condition $\varepsilon_{if}'\leq w_{_{\text{D}}}$ we have $2W<1$ and
the expansion~(\ref{eq:resratsol5}) is used. The two strong peaks
represent the recoilless formation process, without phonon
excitations.  The delta function profile of every peak is shown as
a~rectangle with a~height equal to the formation rate strength divided
by the total decay width ($\approx 0.8\times 10^{-3}$~meV).  The
strength defined as the value of the factor standing before
$\delta(\omega)$ in the expansion~(\ref{eq:resratsol5}), is equal to
0.1061~eV$\cdot\mu$s$^{-1}$ for the resonance $K_i=0 \to K_f=1$ in
solid ortho-D$_2$. The transition $K_i=1 \to K_f=2$ in para-D$_2$
gives 0.07544~eV$\cdot\mu$s$^{-1}$ as the resonance strength.  Higher
resonance energies involve many multi-phonon terms and therefore we
use the asymptotic form~(\ref{eq:resp_asym}) of $S_i$ for
$\varepsilon_{if}'>w_{_{\text{D}}}$. All formation rates presented in the
figures are normalized to the liquid hydrogen density $N_0=4.25\times
10^{22}$~atoms/cm$^3$.

Though in Monte Carlo simulations, involving energy-dependent rates of
different processes, the ``absolute'' formation
rates~$\lambda^F_{K_i}(\varepsilon)$ should be used, it is convenient
to introduce an effective formation
rate~$\bar{\lambda}^F_{K_i}(\varepsilon)$ which leads to the nuclear
$dd$ fusion in $[(dd\mu)dee]$ complex. Back decay of the complex to
the $d\mu+$D$_2$ system, characterized by the quantum numbers $K_i'$
and $F'$, strongly influences the fusion process because the
back-decay rates are comparable with the effective fusion rate
$\bar{\lambda}_f\approx$~374~$\mu$s$^{-1}$~\cite{mens87}. Since in
a~solid target rotational deexcitation of the asymmetric complex is
much faster than back decay and fusion, it is assumed that back decay
takes place only from the state~$K_f=0$. The effective formation rate
is then defined by the following formula
\begin{equation}
  \label{eq:form_eff}
  \bar{\lambda}^F_{K_i}(\varepsilon) = \sum_{K_f,S}
  \lambda^{SF}_{\nu_i K_i \nu_f K_f}(\varepsilon) \, 
  {\mathcal P}^{\text{fus}}_S \,, \qquad \nu_i=0, 
  \quad \nu_f=7  \,,
\end{equation}
where the fusion fraction ${\mathcal P}^{\text{fus}}_S$ is given by
\begin{equation}
  \label{eq:fus_frac}
  {\mathcal P}^{\text{fus}}_S = 
  \frac{\bar{\lambda}_f}{\Gamma^S} \,, \qquad
  \Gamma^S = \bar{\lambda}_f+\sum_{F'} \Gamma^{SF'} \,, \qquad
  \Gamma^{SF'} = \sum_{K_i',K_f=0} 
  \Gamma^{SF'}_{\nu_fK_f,\nu_iK_i'}   \,.
\end{equation}

Since the frequency of lattice vibrations 
($\sim w_{_{\text{D}}}/\hbar\sim 10^{7}$~$\mu$s$^{-1}$) is many orders 
of magnitude greater than the back-decay and fusion rates, energetic
phonons created during the $dd\mu$ formation process are dissipated.
At 3~K the number of phonons with energies
 $w\gtrsim k_{\text{B}}T\approx$~0.26~meV is strongly suppressed by the
Bose factor~$n_{_{\text{B}}}(w)$. Therefore back decay with phonon
annihilation at $T\ll\Theta_{\text{D}}$ is negligible. In particular,
the phonon channel of decay of $dd\mu$, formed from $d\mu$ state
$F=\tfrac{3}{2}$ due to the subthreshold resonances, is closed because
this would require an annihilation of a~phonon with energy of a~few
meV. In this case back decay is connected with the spin-flip
transition to $F'=\tfrac{1}{2}$.  Since the corresponding energy
release of a~few tens of meV is much greater than the Debye energy
($\Delta E^{\text{hfs}}\gg w_D$), the $dd\mu$ decay rate is dominated
by contributions from simultaneous phonon creation processes.

After integration of formula~(\ref{eq:elwidthsol1}) over direction
of vector~$\vec{k}$ one obtains
\begin{equation}
  \label{eq:width_integ}
  \Gamma^{SF'}_{\nu_fK_f,\nu_iK_i} = \frac{A_{if} }{\pi} 
  \int_{0}^{\infty} dk \, k^2 \, \lvert V_{if}(\varepsilon)\rvert^2
   \, \widetilde{{\mathcal S}}_i(k^2,\omega') \,,
\end{equation}
and then substitution of expansion~(\ref{eq:resp2}) and integration of
the recoilless term lead to
\begin{equation}
  \begin{split}
    \label{eq:width_expan}
    \Gamma^{SF'}_{\nu_fK_f,\nu_iK_i} = \frac{A_{if} }{\pi}
    \Bigg[ M &\widetilde{k}_{if} \lvert V_{if}
    (\tilde{\varepsilon}'_{if})\rvert^2 \exp(-2\widetilde{W}_{if})\\
    +& \sum_{n=1}^{\infty} \int_{0}^{\infty} dk \, k^2 \,
    \lvert V_{if}(\varepsilon)\rvert^2 \exp(-2\widetilde{W}) \, 
    g_n(\omega',T)\, \frac{(2\widetilde{W})^n}{n !} \Bigg] \,,
\end{split}
\end{equation}
where
\begin{equation}
  \label{eq:2W_cplx}
  2\widetilde{W} = \frac{k^2}{2M_C}\, \gamma(\infty) \,, \qquad 
  2\widetilde{W}_{if} = 2\widetilde{W}(\widetilde{k}_{if}) \,, \qquad 
  \widetilde{k}_{if} = \sqrt{\,2 M \tilde{\varepsilon}'_{if}} \,.
\end{equation}
It is assumed in the formula above that the phonon energy spectrum of
solid deuterium containing $[(dd\mu)dee]$ is similar to that of a~pure
deuterium lattice. The problem of lattice dynamics of a~quantum solid
deuterium crystal containing a~small admixture of a~heavier isotope
has not been considered yet in literature, at least to the knowledge
of the authors. However, this approximation is reasonable since the
Debye temperatures of solid hydrogen and deuterium at 3~K are very
similar~\cite{silv80}, independently of the mass difference of these
isotopes. Therefore it is assumed that during the $dd\mu$ lifetime the
mean kinetic energy~$\widetilde{{\mathcal E}}_T$ of the complex
reaches the energy~${\mathcal E}_T$ characterizing a~pure deuterium
solid.  Thus the resonance energy shift~(\ref{eq:eres_cplx}) is
approximated by
\begin{equation}
  \label{eq:eres_cplx_ap}
  \Delta\tilde{\varepsilon}_{if} \approx -\left(M_C / 
  M_{\text{mol}} - 1\right)\, {\mathcal E}_T \approx -2.77 
  \text{ meV} \,,
\end{equation}
which gives $\tilde{\varepsilon}'_{if}=\varepsilon_{if}-2.77$~meV.
   
\begin{figure}[htbp]
  \begin{center}
    \epsfig{file=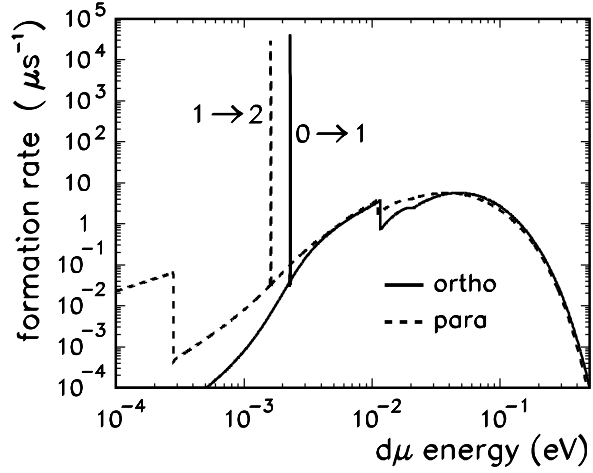, width=8.5cm}    
    \caption{Effective formation rate 
      $\bar{\lambda}^F_{K_i}(\varepsilon)$ for 
      $F=\tfrac{3}{2}$ in 3~K solid ortho-D$_2$ and para-D$_2$. 
      The labels ``$1\to 2$'' and  ``$0\to 1$'' denote the rotational 
      transition $K_i\to K_f$ corresponding to the lowest non-phonon 
      processes.}
    \label{fig:eddm3}
  \end{center}
\end{figure}

The effective formation rates in 3~K solid deuterium for
$F=\tfrac{3}{2}$ are shown in Fig.~\ref{fig:eddm3}.  The phonon part
of the rates below a~few meV is about two orders of magnitude lower
than the average rate of 2.7~$\mu$s$^{-1}$ derived from the
experiment~\cite{know96,know97}. This means that at 
$\varepsilon\ll w_D$ the phonon contribution to the total resonant 
formation rate is even smaller than the non resonant $dd\mu$ formation
rate $\lambda_{\text{nr}}\approx$~0.44~$\mu$s$^{-1}$~\cite{scri93},
and that the estimation of the phonon contribution given in
Ref.~\cite{mens96} is strongly overestimated. Therefore, the
experimental results can only be explained by resonant $dd\mu$
formation at energies $\varepsilon\gtrsim 1$~meV, where the rate
exceeds significantly the value of~1~$\mu$s$^{-1}$. A~cusp at 0.3~meV
in para-D$_2$ is due to the formation with simultaneous one-phonon
creation, connected with the subthreshold resonance $K_i=1\to K_f=0$.
This implies a~significant difference between the resonant formation
in ortho-D$_2$ and para-D$_2$ below 1~meV. However, this difference is
difficult to measure because of a~broad distribution of $d\mu$ energy.
Note that a~similar subthreshold phonon effect in the case of resonant
$dt\mu$ formation in solid deuterium has been discussed
in~Ref~\cite{fuku93}.

In the solid target the fusion fraction 
${\mathcal P}^{\text{fus}}_S\approx$~0.3 and the total resonance 
width $\Gamma^S\approx 0.8\times 10^{-3}$~meV for both 
$S=\tfrac{1}{2}$ and $S=\tfrac{3}{2}$. The back-decay rate~$\Gamma^{SF'}$ 
from $S=\tfrac{1}{2}$ to $F'=\tfrac{1}{2}$ equals about
843~$\mu$s$^{-1}$. Decay $S=\tfrac{1}{2} \to F'=\tfrac{1}{2}$ is
impossible. In the case of $S=\tfrac{3}{2}$ we have obtained
$\Gamma^{SF'}\approx$~281~$\mu$s$^{-1}$ for $F'=\tfrac{1}{2}$ and
$\Gamma^{SF'}\approx$~610~$\mu$s$^{-1}$ for $F'=\tfrac{3}{2}$. Phonon
creation processes give dominant contributions to the back-decay rates,
e.g., the non-phonon part of $\Gamma^{SF'}$, given by the first term
of expansion~(\ref{eq:width_expan}), equals 169~$\mu$s$^{-1}$.
Therefore the $d\mu$ energy spectrum, after back decay in the solid, 
is not discrete.

\begin{figure}[htbp]
  \begin{center}
    \epsfig{file=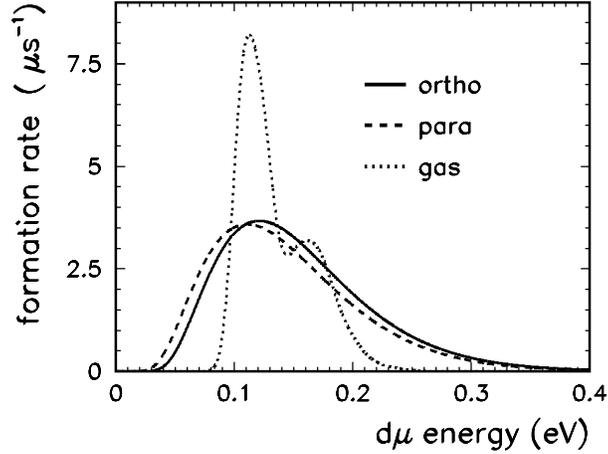, width=8.5cm}     
    \caption{Effective formation rate 
      $\bar{\lambda}^F_{K_i}(\varepsilon)$ for 
      $F=\tfrac{1}{2}$ in 3~K solid  ortho-D$_2$ and para-D$_2$. 
      The label ``gas'' denotes the curve obtained for 3~K gaseous 
      deuterium ($K_i=0$), using the asymptotic 
      formula~(\ref{eq:resp_asym}) for the response
      function~${\mathcal S}_i$ with $T_{\text{eff}}=3$~K.}
    \label{fig:eddm1}
  \end{center}
\end{figure}

In Fig.~\ref{fig:eddm1} the effective rates in solid deuterium for
$F=\tfrac{1}{2}$ are presented. For the sake of comparison the
formation rate for 3~K ortho-D$_2$ gas is also plotted. The ``gas''
curve has been calculated using the asymptotic
formula~(\ref{eq:resp_asym}) for ${\mathcal S}_i$ with 
$T_{\text{eff}}=3$~K.This figure shows that in a~real solid deuterium
target the rates are smeared much more than in a~gas target with the
same temperature, because of the zero-point vibrations. Therefore even
at relatively high $d\mu$ energies of some 0.1~eV one should not
neglect the solid effects and use the formation rates calculated for
a~3~K Maxwellian gas.


\section{Monte Carlo calculations}
\label{sec:Monte_Carlo_form}

\begin{figure}[htbp]
  \begin{center}
    \epsfig{file=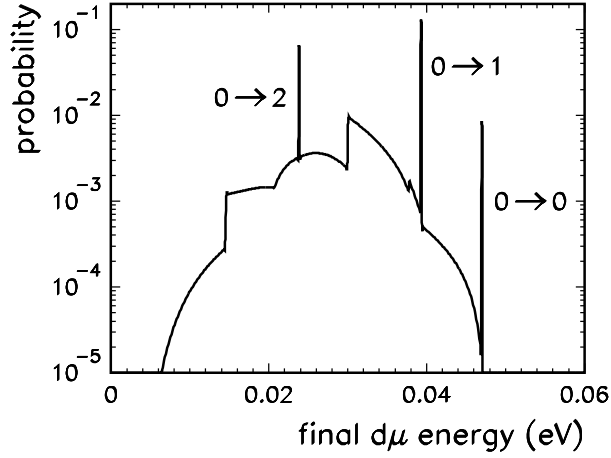, width=8.5cm}
    \caption{Distribution of final $d\mu$ energy after $dd\mu$ back
      decay from $S=\tfrac{1}{2}, K_f=0$ to $F'=\tfrac{1}{2}$, 
      $K_i'=0,1,2$. The three peaks describe the rotational
      transitions without a~simultaneous phonon excitation.}
    \label{fig:ebck}
  \end{center}
\end{figure}

The calculated energy-dependent $dd\mu$ formation rates have been
applied in our Monte Carlo simulations of $\mu$CF in 3~K solid
deuterium targets. The final $d\mu$ energy distribution after back
decay, including simultaneous phonon creation processes, has been
determined through a~numerical integration of
Eq.~(\ref{eq:width_expan}). The calculated distribution is shown in
Fig.~\ref{fig:ebck} for $S=\tfrac{1}{2}$, $K_f=0$ and
$F'=\tfrac{1}{2}$. The rotational transitions to $K_i'=0, 1, 2$ with
no phonon creation are seen as the delta peaks.  The continuous energy
spectrum describes phonon creation contribution to $d\mu$ energy. Note
that, opposite to $dd\mu$ formation rates, this phonon contribution
(for a~given rotational transition peak) extends towards lower
energies. The average $d\mu$ energy after $dd\mu$ back decay equals
about 30~meV, for the presented spectrum.

The $dd$ fusion neutron and proton spectra depend on the time
evolution of $d\mu$ energy. This energy is determined by differential
cross sections of different scattering processes of $d\mu$ atoms in
a~given solid target, including elastic scattering, rovibrational
transitions, spin-flip reactions and phonon processes. The scattering
cross sections in a~solid are calculated using the Van Hove method.
Some results of such calculations for $d\mu$ atoms in fcc solid
deuterium have been presented and discussed in Ref.~\cite{adam99}.
The incoherent processes, such as spin-flip or rovibrational
transitions, are described by the self pair correlation function
$G_s(\vec{r},t)$ defined by Eq.~(\ref{eq:selfcor}). The Bragg
scattering and coherent phonon scattering are connected with a~pair
correlation function~$G(\vec{r},t)$~\cite{vanh54}.

\begin{figure}[htbp]
  \begin{center}
    \epsfig{file=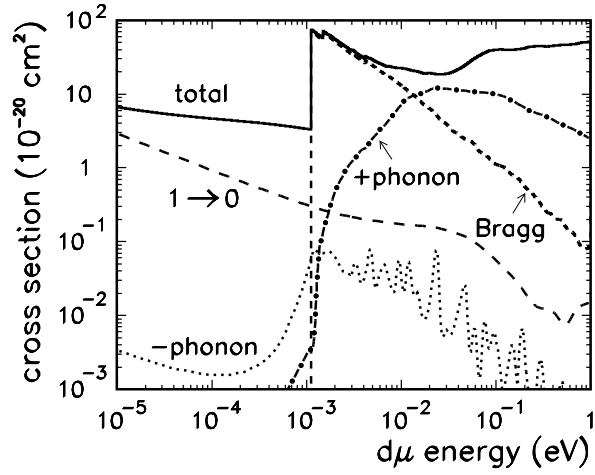, width=8.5cm}
    \caption{Total cross section for $d\mu(F=\tfrac{3}{2})$ scattering
      in statistical mixture of solid ortho-D$_2$ and para-D$_2$. The 
      label ``$1\to 0$'' denotes the rotational deexcitation 
      $K=0\to 1$ of a~target D$_2$ molecule. The curves ``$-$phonon'' 
      and ``+phonon'' stand for $d\mu$ scattering with phonon 
      annihilation and creation, respectively. The Bragg cross section
      is calculated for the fcc polycrystalline lattice.}
    \label{fig:xfddd22}
  \end{center}
\end{figure}

In Fig.~\ref{fig:xfddd22} is shown the total cross section for
$d\mu(F=\tfrac{3}{2})$ scattering in the statistical mixture of 3~K
solid ortho-D$_2$ and para-D$_2$. Bragg scattering, with the Bragg
cutoff at $\varepsilon_{_{\text{B}}}=$~1.1~meV, and incoherent elastic
scattering do not change $d\mu$ energy because of the very large mass
of the considered solid target. Below 1.7~meV the $d\mu$ atom is
effectively accelerated, mainly due to the rotational deexcitation of
para-D$_2$ molecules~\cite{guri99,adam99}. This transition is enabled
by muon exchange between deuterons in $d\mu+$D$_2$ scattering. The
curve ``$0\to 1$'' in Fig.~\ref{fig:xfddd22}, describing the
rotational deexcitation, includes contributions from simultaneous
incoherent phonon processes.  This cross section at
$\varepsilon=$~2.5~meV equals 0.22$\times 10^{-20}$~cm$^2$, which is
about three times less (taking into account the statistical factor of
1/3 for $K=1$ states) than the estimation given in
paper~\cite{guri99}.  Phonon annihilation is a~much weaker $d\mu$
acceleration mechanism than the rotational deexcitation.

Since the coherent amplitude for $d\mu$ elastic scattering on a~single
D$_2$ molecule is greater by two orders of magnitude than the
incoherent amplitude, the coherent processes involving conservation of
momentum dominate low energy $d\mu$ scattering in solid deuterium. It
is especially important below a~few meV, where the shapes of coherent
and incoherent cross sections differ strongly. The small phonon
creation cross section below 1.1~meV, leading to $d\mu$ energy loss,
is due to the incoherent amplitude. Coherent phonon creation is
impossible below~$\varepsilon_{_{\text{B}}}$. This limit is obtained in
the case of coherent one-phonon creation process, for the total
momentum conservation involving the smallest (non-zero) inverse
lattice vector~$\vec{\tau}$, which also fixes the position of the
first peak of the Bragg scattering 
at~$\varepsilon_{_{\text{B}}}=$~1.1~meV. For $\vec{\tau}=\vec{0}$
one-phonon creation is possible only if the $d\mu$ velocity is not
lower than the sound velocity in the crystal, which is well-known in
neutron physics.  According to Ref.~\cite{soue86} the mean sound
velocity in solid deuterium equals about 1.2$\times 10^5$~cm/s and
this corresponds to $d\mu$ energy of 15~meV.  Therefore, neglecting
the inverse lattice contribution to the one-phonon creation cross
section in Ref.~\cite{guri99} leads to the severe underestimation of
$d\mu$ slowing down at lowest energies and subsequent overestimation
of $d\mu$ kinetic energy.

Above 1.7~meV phonon creation already prevails over all acceleration
processes. However, the effective deceleration rate below
$w_{_{\text{D}}}$ is strongly suppressed by the dominating Bragg elastic
scattering. At energies above some 10~meV subsequent rotational and
then vibrational excitations of D$_2$ molecules become important and
they provide a very fast mechanism of $d\mu$ deceleration at higher
energies.

\begin{figure}[htbp]
  \begin{center}
    \epsfig{file=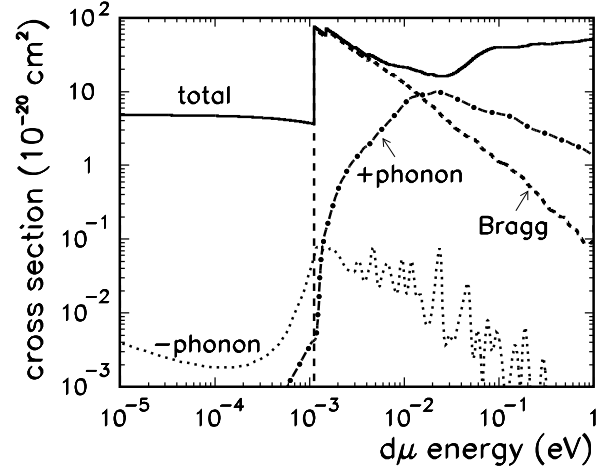, width=8.5cm}
    \caption{Cross section for $d\mu(F=\tfrac{3}{2})$ scattering in 
      solid ortho-D$_2$. The labels are identical to those in
      Fig.~\ref{fig:xfddd22}.}
    \label{fig:xfddd22k0}
  \end{center}
\end{figure}

The total cross section for $d\mu(F=\tfrac{3}{2})$ scattering in
a~pure 3~K ortho-D$_2$ target (see Fig.~\ref{fig:xfddd22k0}) is
quite similar to that shown in Fig.~\ref{fig:xfddd22}. A~significant
difference is the lack of rotational deexcitation. Therefore phonon
annihilation is the only, and weak, acceleration mechanism. It
dominates the inelastic cross section below 1.4~meV.

\begin{figure}[htbp]
  \begin{center}
    \epsfig{file=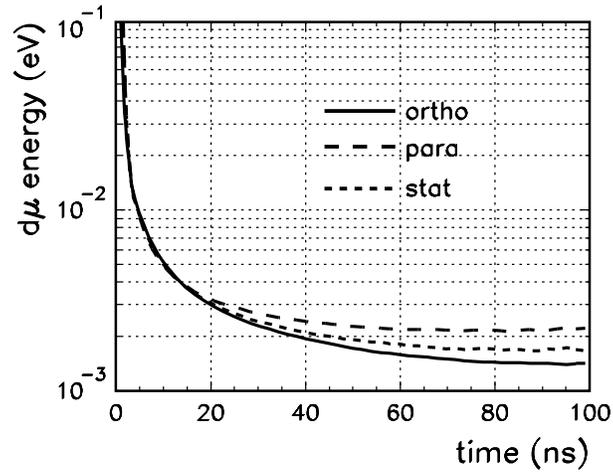, width=8.5cm}
    \caption{Calculated time evolution of average $d\mu$
      energy~$\varepsilon_{\text{avg}}$ for $F=\tfrac{3}{2}$ in 3~K
      solid ortho-D$_2$, para-D$_2$, and their statistical mixture
      (stat). A~Maxwell distribution of $d\mu$ initial energy, with
      mean energy of 1~eV, has been assumed.}
    \label{fig:avendm3}
  \end{center}
\end{figure}

Fig.~\ref{fig:avendm3} presents the time evolution of average
$d\mu(F=\tfrac{3}{2})$ atom energy~$\varepsilon_{\text{avg}}$,
obtained from our Monte Carlo calculations. It is assumed that the
target is infinite and that $d\mu$ atoms have initially a~Maxwellian
energy distribution with a~mean energy of~1~eV. A~statistical initial
population of $d\mu$ total spin is used and the theoretical
non-resonant part of the total spin-flip rate
$\lambda_{\frac{3}{2},\frac{1}{2}}$ is multiplied by a~single scaling
factor of~0.4, in order to keep agreement with the experimental
values~\cite{peti99,voro99} of the spin-flip rate. The calculations
have been performed for ortho-D$_2$, para-D$_2$ and their statistical
mixture (stat). One can see that $d\mu$ mean energy of 10~meV is
reached already after 5~ns. Then, below the Debye energy, deceleration
become very slow. The lowest value of $\varepsilon_{\text{avg}}$ is
determined by the intersection point of the cross sections of the
acceleration processes and phonon creation process. In the case of
a~statistical mixture $\varepsilon_{\text{avg}}\approx$~1.7~meV, for
$K=0$ we have $\varepsilon_{\text{avg}}\approx$~1.4~meV. Finally, for
pure para-D$_2$, with a~contribution to the total cross section from
the rotational transition $K=1\to 0$ three times greater than that
shown in Fig.~\ref{fig:xfddd22}, 
$\varepsilon_{\text{avg}}\approx$~2.2~meV. 
Thus, $d\mu$ atoms are never thermalized
and their energy is significantly greater than 1~meV. For para-D$_2$
the mean energy is always greater than the energy of the lowest
resonance peak $\varepsilon_{if}'=$~1.6~meV.  However, even if
$\varepsilon_{\text{avg}}$ is smaller than~$\varepsilon_{if}'$,
a~significant part of $d\mu$ atoms has energy
$\varepsilon\geq\varepsilon_{if}'$ because of a~large admixture of hot
$d\mu$ atoms at $t=0$~\cite{mark94,abbo97} and slow deceleration below
10~meV.

\begin{figure}[htbp]
  \begin{center}
    \epsfig{file=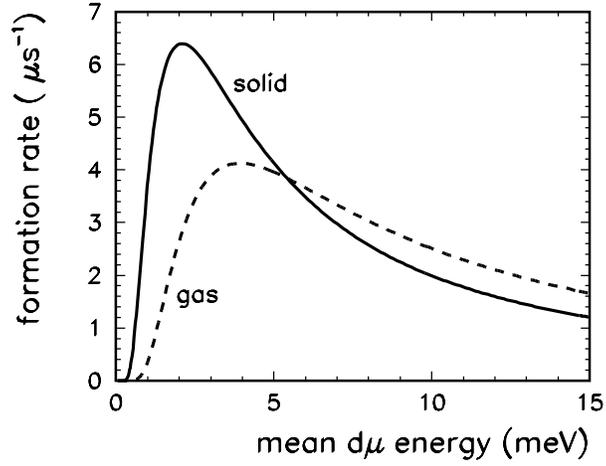, width=8.5cm}    
    \caption{The effective resonant $dd\mu$ formation rate as 
      a~function of mean CMS energy~$\varepsilon_{\text{avg}}$ of 
      $d\mu(F=\tfrac{3}{2})$ atom for gas and solid deuterium
      targets. A~steady Maxwell distribution of $d\mu$ energy is
      assumed for  a~given~$\varepsilon_{\text{avg}}$. The 
      contributions from the two lowest resonant peaks to the
      formation rate are taken into account.}
    \label{fig:vesav}
  \end{center}
\end{figure}

\begin{figure}[htbp]
  \begin{center}
    \epsfig{file=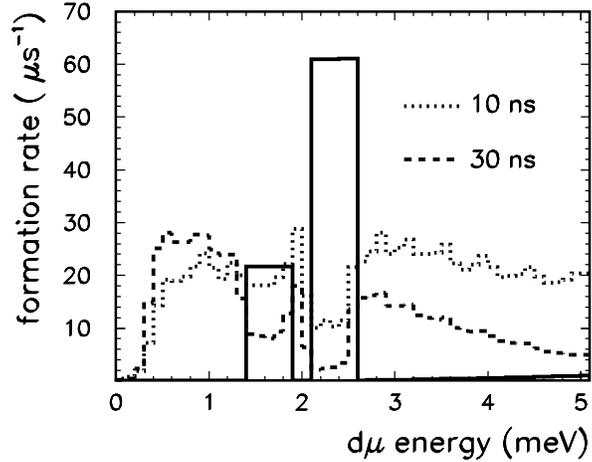, width=8.5cm}        
    \caption{Resonant $dd\mu$ formation rate for $F=\tfrac{3}{2}$ in 
      the statistical mixture of ortho-D$_2$ and para-D$_2$ for the
      resonance peak width $\Gamma^S=$~0.5~meV. Monte Carlo
      distribution of $d\mu$ energy at $t$=10~ns and $t$=30~ns after
      the muon stop is plotted (in arbitrary units). }
    \label{fig:eddmdistat3}
  \end{center}
\end{figure}

Since at energies of a~few meV the lowest delta peaks are dominant in
the resonant formation, their contributions to the mean effective
formation rate are shown in Fig.~\ref{fig:vesav} for gas and solid
deuterium (stat) targets, assuming steady Maxwell distributions of
$d\mu(F=\tfrac{3}{2})$ energy, with
different~$\varepsilon_{\text{avg}}$.  The maximum average rate of
about 6~$\mu$s$^{-1}$ in the solid is due to the resonance energy
shift of $-$1.8~meV. The experimental result of 3~$\mu$s$^{-1}$ can be
explained because $\varepsilon_{\text{avg}}$ is greater than 1~meV.
However, in order to obtain large fusion neutron and proton yields
through resonant $dd\mu$ formation, the width $\Gamma^S$ of the
resonance peaks in solid can not be too narrow. The peak resonant
rates of a~few 10$^4 \mu$s$^{-1}$ have been obtained assuming the
discrete values of the rovibrational D$_2$ energies in solid deuterium
and $\Gamma^S\sim 10^{-3}$~meV. These resonant rates are many orders
of magnitude greater than the inelastic scattering rate
$\sim$~10~$\mu$s$^{-1}$. In such a~case $d\mu$ atoms are very quickly
(compared to $d\mu(F=\tfrac{3}{2})$ lifetime) removed from the regions
of resonance peaks and the contribution of the recoilless resonances
to the neutron yield is negligible. The Monte Carlo simulations have
shown that the neutron yield from the phonon part of the resonant
rates gives only some 10\% of the yield observed in the experiments.
In result, the calculated time spectra, obtained for the small
$\Gamma^S$, are dominated by weak non-resonant $dd\mu$ formation,
which disagrees with the experimental data. Therefore, we have
investigated influence of a~broadening of the non-phonon resonant
peaks, due to the presence of molecular rovibrational bands in solid,
discussed in Ref.~\cite{vank83}. Since in the literature there is no
information concerning the profile of such bands, we have assumed
a~rectangular shape of the resonance peaks. The resonance strengths
have been fixed and their widths have been changed in the limits
0.001--1~meV. It turns out that good Monte Carlo results are obtained
for $\Gamma^S\approx$~0.5~meV, which is consistent with the rotational
bandwidths of about 1~meV reported in Ref.~\cite{vank83}. This gives
the resonant formation rate of 294~$\mu$s$^{-1}$ for the recoilless
peak in ortho-D$_2$\@, and respectively 214~$\mu$s$^{-1}$ in
para-D$_2$.  In Fig.~\ref{fig:eddmdistat3} one sees the resonant
formation rate at lowest energies for $\Gamma^S=$~0.5~meV and for the
statistical mixture of ortho- and para-states. Also shown is the Monte
Carlo distribution of $d\mu(F=\tfrac{3}{2})$ energy, calculated for
times $t=$~10~ns and $t=$~30~ns. The Maxwell distribution of initial
$d\mu$ energy, with~$\varepsilon_{\text{avg}}=$~1~eV, has been
assumed. Two minima in the $d\mu$ energy distribution appear quickly
at the positions of the resonance peaks since the respective $dd\mu$
formation rates are comparable with the total inelastic scattering
rate of about 30~$\mu$s$^{-1}$.

\begin{figure}[htbp]
  \begin{center}
    \epsfig{file=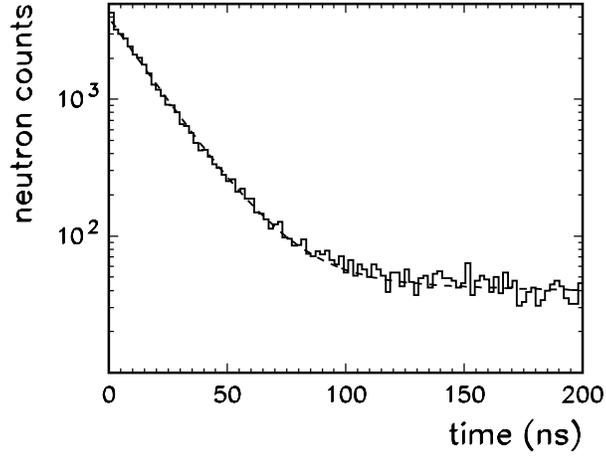, width=8.5cm}
    \caption{The Monte Carlo fusion neutron spectrum for the statistical
      mixture of 3~K solid ortho-D$_2$ and para-D$_2$ (solid line). The
      dashed line represents the spectrum obtained using an analytical
      steady state kinetics model with 
      $\bar{\lambda}^{3/2}_{stat}=$~3~$\mu$s$^{-1}$.
      The initial $d\mu$ energy is given
      by a~Maxwell distribution with mean energy of~1~eV. The
      width~$\Gamma ^S$ of the non-phonon resonances is fixed at
      0.5~meV. A~3.2$\times 10^{-6}$ concentration of nitrogen is
      included.}
    \label{fig:npfita}
  \end{center}
\end{figure}

The $dd$ fusion neutron spectrum, calculated assuming the same initial
$d\mu$ energy and resonance profiles, is shown in
Fig.~\ref{fig:npfita}.  A~3.2$\times 10^{-6}$ concentration of
nitrogen is included in order to fit the TRIUMF target conditions.
The solid line plotted in this figure has been calculated using the
steady-state kinetics model with the effective formation rate
$\bar{\lambda}^{3/2}_{stat}=$~3~$\mu$s$^{-1}$ and total
spin-flip rate
$\lambda_{\frac{3}{2},\frac{1}{2}}=$~36~$\mu$s$^{-1}$ taken from
the fits to the experimental data~\cite{know97}. The slope of the
spectrum at $t\lesssim$~80 ns is determined by the rates
$\bar{\lambda}^{3/2}_{stat}$, $\lambda_{\frac{3}{2},\frac{1}{2}}$, and
$d\mu$ scattering rate which also changes the population of
$d\mu(F=\tfrac{3}{2})$ atoms in the vicinity of the resonant peaks.
The steady-state kinetics model does not include the process of $d\mu$
deceleration. Therefore, fits using this model could entangle the
deceleration rate with the formation and spin-flip rates. The mean
formation rate, calculated directly in the Monte Carlo runs, is
a~function of time, and it stays at the level of 1--3~$\mu$s$^{-1}$.
The spectrum slope at large times $t\gtrsim$~100~ns, when
$d\mu(F=\tfrac{3}{2})$ atoms practically disappear, are due to the
nonresonant $dd\mu$ formation from~$F=\tfrac{1}{2}$ and to the muon
transfer to nitrogen contamination.

The shape of the time spectra practically does not change when the
mean energy $\varepsilon_{\text{avg}}$ of the initial single Maxwell
distribution varies in the limits 0.01--1~eV. On the other hand, the
spectra change strongly if a~significant part of $d\mu$ atoms at $t=0$
has energy smaller than the energy of the lowest resonant peak, which
can be observed using a~more complicated (e.g.\ two-Maxwell
distribution). Assuming that $\Gamma^S$ is greater than 0.5~meV we
obtain results which begin to differ significantly from the analytical
curve calculated with the experimental parameters. In particular, the
ratios of neutron yields from the short and large times begin to
disagree. Fits of the calculated spectra to the experimental data
would enable a~better determining of~$\Gamma^S$ and a~shape of the
initial $d\mu$ energy. However, this is not the purpose of this work.
A~qualitative comparison of Monte Carlo spectra with the experimental
data has already been performed in article~\cite{know97}. In this case
good fits were not obtained since at that time the resonant $dd\mu$
formation rates in solid D$_2$ and $d\mu$ scattering rates including
coherent effects in the solid were not yet available.

\begin{figure}[htbp]
  \begin{center}
    \epsfig{file=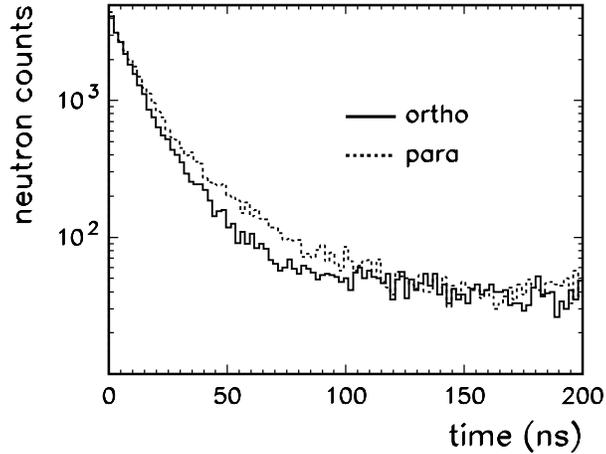, width=8.5cm}    
    \caption{Calculated neutron spectra from 3~K solid
      ortho-D$_2$ and para-D$_2$. The Maxwell distribution of initial
      $d\mu$ energy with $\varepsilon_{\text{avg}}=1$~eV and
      $\Gamma^S=$0.5~meV have been assumed for the both targets.}
    \label{fig:neutronl}
  \end{center}
\end{figure}

Our calculations show that strong resonant $dd\mu$ formation takes
place both in ortho-D$_2$ and para-D$_2$.  There are certain
differences between the neutron time spectra from these targets (see
Fig.~\ref{fig:neutronl}), caused by the different positions and
strengths of the lowest resonance peaks. Also $d\mu$ slowing down
process differs slightly in the two cases. The neutron yield at larger
times is smaller for ortho-D$_2$ since in this case the resonance peak
is placed at higher energy of 2.3~meV. Therefore, $d\mu$ atoms are
removed faster from the peak compared to the situation in para-D$_2$,
where the resonance is observed at 1.6~meV.  A~greater mean $d\mu$
energy in para-D$_2$ (cf.\ Fig.~\ref{fig:avendm3}) leads also to
a~stronger overlap of the resonance peak and $d\mu$ energy
distribution at $t\gtrsim 20$~ns.  However, the differences between
the spectra can be clearly seen only in high-statistics experiments.


\section{Conclusions}
\label{sec:concl_form}

The methods used for description of resonant neutron and $\gamma$-ray
absorption in condensed matter have been directly applied for
calculation of resonant $dd\mu$ formation and back-decay rates in
condensed deuterium targets. These rates are expressed in terms of the
Van Hove single-particle function, which depends on properties of
a~given target. In particular, we have derived the analytical formulas
for the rate in the case of resonant $dd\mu$ formation in a~harmonic
solid deuterium. The calculations show great differences between
resonant $dd\mu$ formation in 3~K solid deuterium and in 3~K D$_2$
gas. In solid, the formation at a~few meV, which determines the
experimental results, is dominated by presence of the strong
recoilless resonant peaks. On the other hand, the formation with
simultaneous phonon creation is important above the Debye energy. The
resonance profiles in the solid at higher energies are similar to that
in D$_2$ gas, but with the effective temperature equal to~40~K. This
temperature is determined by the energy of zero-point vibration of
D$_2$ molecule in the lattice. Phonon creation is always important
in the case of $dd\mu$ back decay because it is connected with energy
release of a~few tens meV, which is much greater than the Debye
energy.

A~condition $T/\Theta_{_D}\ll 1$ is fulfilled for any solid deuterium
target at low pressure. Therefore, the parameters determining solid
state effects (Debye-Waller factor, mean energy of D$_2$ vibration in
solid) weakly depend on target temperature~$T$. They are expressed in
terms of the Debye energy~$w_{_{\text{D}}}$ which does not significantly
change with the varying solid temperature~$T$. In result, the resonant
$dd\mu$ formation rates in solid deuterium for different~$T$ are very
similar and one may expect that the average formation rates, derived
from measurements performed at different temperatures, will also be
very close. This is confirmed by the results of experiments carried
out at TRIUMF and at JINR.

The structure of a~solid deuterium target depends on its temperature
and history.  Targets maintained at $T\gtrsim$~4~K have the hcp
structure~\cite{silv80}. Though our calculations have been performed
for fcc crystals, the obtained results are also good approximations of
the resonant rates in hcp polycrystals since the Debye temperature and
nearest neighbor distance are similar for these two lattices. In
general, the formulas derived in this paper can be used in a~wide
range of target temperature and density, with appropriate experimental
values of the Debye temperature and lattice constant taken into
account.

The Monte Carlo calculations show that $d\mu$ deceleration below the
Debye energy is very slow and that mean energy of
$d\mu(F=\tfrac{3}{2})$ atom is always significantly greater than
1~meV. The energy distribution of $d\mu$'s during their lifetime is
very broad (at least a~few meV), therefore a~strong overlap of this
distribution and lowest resonance peaks takes place, leading to
a~large mean $dd\mu$ formation rate in solid deuterium. However,
explanation of the experiments is possible only if the broadening of
rovibrational molecular levels in solid is taken into account. We
obtained reasonable results assuming that the strengths of the
recoilless resonant peaks are constant and that the rotational bands
increase the resonance peak width to 0.5~meV.  Note that, according to
Ref.~\cite{vank83}, high pressures lead to a~greater broadening and
even to a~mixing of rotational states. This could complicate
a~comparison of theory and high-pressure experiments.  The phonon part
of the resonant rate give only about 10\% contribution to the
calculated neutron time spectra.

The $dd$ fusion neutron spectra calculated for ortho-D$_2$ and
para-D$_2$ solid targets are quite similar. Small differences between
the spectra are due to the different energies and strengths of the
lowest resonant peaks, and to a~slightly higher mean $d\mu$ energy in
para-D$_2$. These differences can be clearly seen only in
high-statistics experiments. Our calculations do not confirm a~lack of
strong resonant $dd\mu$ formation in solid ortho-D$_2$, predicted in
the papers~\cite{mens96,guri99}. In order to verify the theory it is
necessary to perform measurements in pure ortho-D$_2$ and para-D$_2$
solid targets under the same conditions.


\acknowledgments 

We wish to thank L.~I.~Ponomarev for stimulationg discussions. We are
grateful to G.~M.~Marshall for a~critical reading of the manuscript.
This work was supported in part through Grant INTAS 97-11032.



%
%

%
%

%
%

%
%

%
%

%
%

%
%

%
%

%
%

%
%

%
%

%
%

%
%

%
%

%
%

%
%



%
\begin{table}[htbp]
  \caption{The lowest resonance energies of $dd\mu$ formation in 
    $d\mu$ scattering from single D$_2$ molecule~($\varepsilon_{if}$) 
    and from 3~K solid deuterium target~($\varepsilon_{if}'$). These
    energies are given in the respective CMS systems.}
  \label{tab:eres}
  \begin{tabular}{r @{.} l r @{.} l c c c c}
    \multicolumn{2}{c}{$\varepsilon_{if}$ (meV)}&
    \multicolumn{2}{c}{$\varepsilon_{if}'$ (meV)}& 
    \multicolumn{1}{c}{$F$}&    
    \multicolumn{1}{c}{$K_i$}&
    \multicolumn{1}{c}{$K_f$}&
    \multicolumn{1}{c}{$S$}    \\
    \hline \hline
    $-$7&218  & $-$9&028 & $\tfrac{3}{2}$ & 1 & 0 & $\tfrac{1}{2}$ \\
    $-$3&667  & $-$5&477 & $\tfrac{3}{2}$ & 1 & 1 & $\tfrac{1}{2}$ \\
       0&5368 & $-$1&272 & $\tfrac{3}{2}$ & 0 & 0 & $\tfrac{1}{2}$ \\
       3&422  &    1&612 & $\tfrac{3}{2}$ & 1 & 2 & $\tfrac{1}{2}$ \\
       4&088  &    2&279 & $\tfrac{3}{2}$ & 0 & 1 & $\tfrac{1}{2}$ \\
      11&18   &    9&368 & $\tfrac{3}{2}$ & 0 & 2 & $\tfrac{1}{2}$ \\
      \hline
      42&10   &    40&30 & $\tfrac{1}{2}$ & 1 & 0 & $\tfrac{1}{2}$ \\
      45&66   &    43&85 & $\tfrac{1}{2}$ & 1 & 1 & $\tfrac{1}{2}$ \\
      49&86   &    48&05 & $\tfrac{1}{2}$ & 0 & 0 & $\tfrac{1}{2}$ \\
      52&74   &    50&94 & $\tfrac{1}{2}$ & 1 & 2 & $\tfrac{1}{2}$ \\
      53&41   &    51&60 & $\tfrac{1}{2}$ & 0 & 1 & $\tfrac{1}{2}$ \\
      60&50   &    58&69 & $\tfrac{1}{2}$ & 0 & 2 & $\tfrac{1}{2}$ \\
      \hline \hline
    \end{tabular}
\end{table}
%


\end{document}